\documentclass[paper]{ieice}
\usepackage[pdftex]{graphicx,xcolor}
\usepackage[fleqn]{amsmath}
\usepackage{newtxtext}
\usepackage[varg]{newtxmath}

\usepackage[hyphens]{url}

\usepackage{algorithm}
\usepackage{algorithmic}

\def\Hline{\noalign{\hrule height 0.4mm}}
 
\field{B}

\title{Image Generative Semantic Communication with Multi-Modal Similarity Estimation for Resource-Limited Networks}

\authorlist{%
 \authorentry[hosonuma@mcl.iis.u-tokyo.ac.jp]{Eri HOSONUMA}{s}{iis}\MembershipNumber{1914616}
 \authorentry{Taku YAMAZAKI}{m}{sit}\MembershipNumber{1205099}
 \authorentry{Takumi MIYOSHI}{m}{sit,iis}\MembershipNumber{9411011}
 \authorentry{Akihito TAYA}{m}{iis}\MembershipNumber{1102117}
 \authorentry{Yuuki NISHIYAMA}{m}{csis}\MembershipNumber{2106193}
 \authorentry{Kaoru SEZAKI}{m}{csis,iis}\MembershipNumber{8417116}
}
\affiliate[iis]{The authors are with Institute of Industrial Science, The University of Tokyo, Meguro-ku, 153-8505, Japan.}
\affiliate[sit]{The authors are with Collage of Systems Engineering and Science, Shibaura Institute of Technology, Saitama-shi, 337-8570, Japan.}
\affiliate[csis]{The authors are with Center for Spatial Information Science, The University of Tokyo, Kashiwa-shi, 277-8568, Japan.}

\received{2015}{1}{1}
\revised{2015}{1}{1}

\begin{document}
\maketitle
\begin{summary}

To reduce network traffic and support environments with limited resources, a method for transmitting images with minimal transmission data is required.
Several machine learning-based image compression methods, which compress the data size of images while maintaining their features, have been proposed.
However, in certain situations, reconstructing only the semantic information of images at the receiver end may be sufficient.
To realize this concept, semantic-information-based communication, called semantic communication, has been proposed, along with an image transmission method using semantic communication.
This method transmits only the semantic information of an image, and the receiver reconstructs it using an image-generation model.
This method utilizes a single type of semantic information for image reconstruction, but reconstructing images similar to the original image using only this information is challenging.
This study proposes a multi-modal image transmission method that leverages various types of semantic information for efficient semantic communication.
The proposed method extracts multi-modal semantic information from an original image and transmits only that to a receiver.
Subsequently, the receiver generates multiple images using an image-generation model and selects an output image based on semantic similarity.
The receiver must select the result based only on the received features; however, evaluating semantic similarity using conventional metrics is challenging.
Therefore, this study explores new metrics to evaluate the similarity between semantic features of images and proposes two scoring procedures for evaluating semantic similarity between images based on multiple semantic features.
The results indicate that the proposed procedures can compare semantic similarities, such as position and composition, between the semantic features of the original and generated images.

\end{summary}

\section{Introduction}
\label{sec:intro}

\begin{figure*}[!t]
  \centering
  \includegraphics[clip,width=0.95\linewidth]{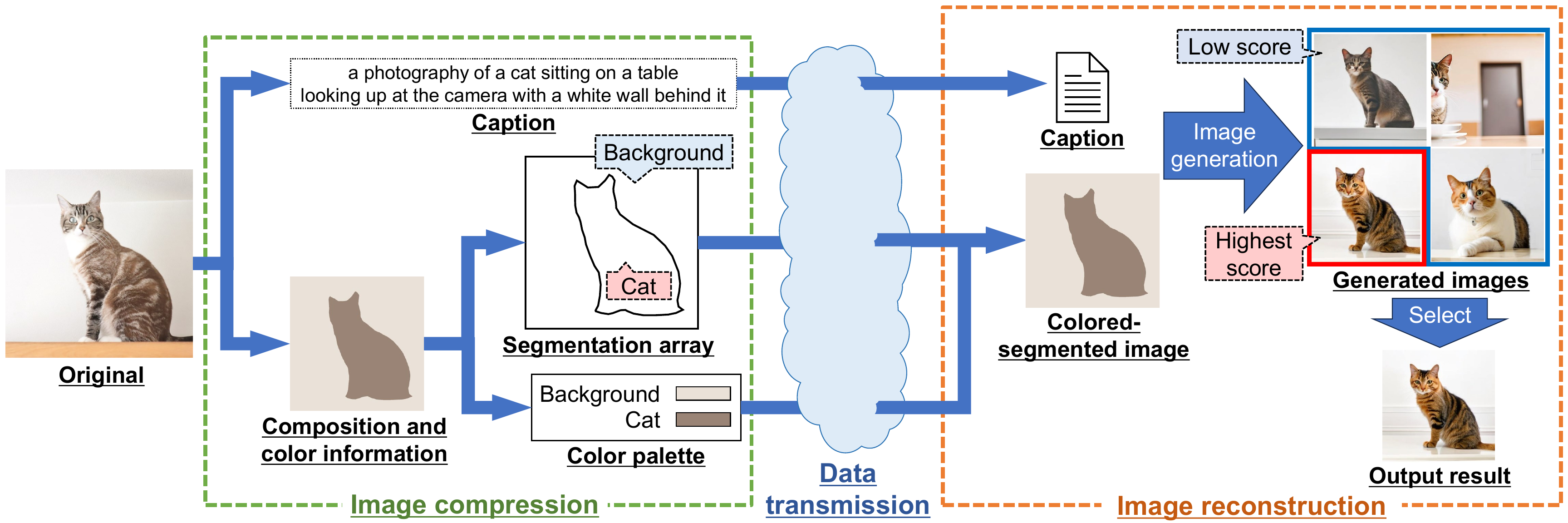}
  \caption{Overview of the proposed image transmission method.
  A transmitter generates multiple features from an image for extracting semantic information and transmits only the features to achieve data reduction.
  A receiver receives the extracted features and reconstructs the image using an image-generation model.}
  \label{fig:overview}
\end{figure*}

\subsection{Background}
\label{sec:background}

The application of photos taken in specific locations for services such as monitoring, detection, and tracking can be utilized in various scenarios.
To realize such applications in multiple locations using lots of devices, sufficient communication resources are required within networks for periodically transmitting images.
However, the following issues may cause a shortage of the communication resources within the networks: (1) Available communication resources are severely limited due to communication media with low data rates, and (2) the resource allocated to each device is limited because of utilizing communication media with high data rates by numerous devices.

The issue (1) may arise in severe environments where mobile networks are unavailable.
Especially, methods for establishing networks in severe environments, such as underwater, mountain, and forest, attract attention in recent years for applications in monitoring, detecting and tracking ~\cite{bib:underwater-1,bib:underwater-2,bib:lpwa}.
For example, underwater communication technologies using acoustic wave and visible light have been proposed for monitoring of underwater environments and surveys of undersea resources in underwater environments~\cite{bib:underwater-1,bib:underwater-2}.
For another example, low power wide area (LPWA) that realizes wide-area communication with low data rates attracts attention for monitoring and detecting in severe terrestrial environments such as mountains or forests~\cite{bib:lpwa,bib:satellite-2}.
These communication technologies have a potential to realize the applications with image transmission in the severe environments.
However, these technologies are difficult to periodically transmit large data such as images at multiple locations because their data rates are lower than those of mobile networks.

The issue (2) may arise in situations where numerous devices need to share abundant communication resources.
Recently, global network traffic is expected to continue increasing due to an increase in the number of devices~\cite{bib:soumu,bib:ericsson,bib:cisco}, and wireless communication technologies such as 5G and 6G rapidly advance to address this issue.
In addition, new communication technologies such as satellite communication have been proposed to cover terrestrial areas without mobile networks and to realize non terrestrial network (NTN)~\cite{bib:satellite-1,bib:satellite-2,bib:satellite-3}.
However, the communication resources allocated to each device become exceedingly limited with an increase in the number of devices, even taking into account developments in these communication technologies.
These trends may make image transmission in terrestrial and non terrestrial environments for monitoring, detecting, and tracking more difficult.
Therefore, image transmission using fewer communication resources has the potential to be utilized in various applications across both urban and remote environments.

\subsection{Semantic Communications}
\label{sec:semantic-communication}

A common approach for reducing the communication cost of image transmission is to use image compression methods~\cite{bib:image-compression,bib:jpeg,bib:jpeg2000}.
Furthermore, machine-learning-based image compression methods that realize high compression rates have been proposed~\cite{bib:learning-base-image-comp-1,bib:learning-base-image-comp-2}.
However, these methods do not consider the transmission process.
In contrast, deep joint source and channel coding (DeepJSCC)~\cite{bib:deep-jscc-1,bib:deep-jscc-2} schemes have been developed to compress and transmit images efficiently by optimizing the joint coding scheme to suit wireless channels.
DeepJSCC encodes images based on the semantic features in the data to be transmitted and realizes image transmission while maintaining their visual information.
Communication technologies, such as DeepJSCC, that focus on the meaning and semantics in communication bits have recently attracted attention as semantic communication~\cite{bib:semantic-commun-1,bib:semantic-commun-2,bib:semantic-commun-3}.

These methods aim to compress and transmit images while maintaining the visual identity of the images. Therefore, these compressed images retain entire of the semantic information in the original images.
However, in environments where communication resources are severely limited, as described in Sec.~\ref{sec:background}, it is difficult to transmit images compressed by these algorithms because the compressed images still have large data sizes.
In addition, transmitting all features of the images is redundant in certain situations. 
For example, when monitoring or detecting vermin in natural environments such as large farms, it will be sufficient to reconstruct types and conditions at a receiver.
In other words, in this situation, background information in images is less important than the information of the target objects.
For another example, when monitoring or detecting specific objects such as vehicles using numerous devices that are allocated limited communication resources in urban environments, background information and facial information of people staying in the urban are not important.
In these situations, it will be sufficient if only specific semantic information contained in the original images, such as presents, types, and compositions of the target objects, is reconstructed by the receiver.

In such situations, if a transmitter can extract specific semantic features from an original image based on the requirements of the receiver, and the receiver can reconstruct the image with the same semantic features in the original one based on these features, the amount of the transmitted data be expected to significantly reduce by transmitting only the extracted features.
Additionally, owing to recent developments in the field of artificial intelligence, image-generation models~\cite{bib:stable-diffusion,bib:dalle,bib:dalle-2} that allow generating images using descriptive information or sketches have become widely available.
Moreover, recent real-time image-generation models~\cite{bib:stream-diffusion} can apply image-generation mechanisms to time-sensitive applications.
Consequently, there is potential for the development of image-generation-based image transmission, which can be considered a type of semantic communication in the future.

As an example of such a kind of communications, the authors of~\cite{bib:t2i-coding} have proposed an image transmission method based on the descriptive information contained in an image.
In this method, the transmitter transmits only the descriptive text information extracted from the original image, and the receiver then reconstructs the image based on the received information through an image-generation model.
Although this method can significantly reduce the communication cost compared to existing image transmission methods, some important semantic features, such as object position and composition, might be lost because it only relies on a single semantic feature in image reconstruction.
Therefore, it is necessary to extract and transmit multi-modal semantic features from the original image to realize image reconstruction with high semantic similarity and satisfy the requirements of the receiver.

\subsection{Image Generation-based Transmission Method with Multi-Modal Semantic Information}
\label{sec:multi-modal-transmission}

To address this issue, we previously proposed a concept of an image transmission method that employs multiple types of semantic information~\cite{bib:ccnc}.
In this method, a transmitter extracts multiple semantic features from an original image, such as composition and color, in addition to captions, and transmits them to a receiver, which reconstructs the image using an image-generation model.
The images generated using multiple semantic features are expected to have a higher similarity to the original images than those generated using only a single feature.
In this method, the receiver needs to evaluate the generated images based on some metrics and select an output result among the images.
However, the specific procedures for the evaluation process and investigating evaluation metrics was not included in~\cite{bib:ccnc}.

In this paper, we propose a highly efficient image transmission method based on the concept proposed in~\cite{bib:ccnc} and employ newly designed evaluation metrics to assess the semantic similarity between the original and generated images.
The proposed method aims to achieve monitoring, detecting and tracking specific objects mentioned in Sec.~\ref{sec:semantic-communication} in environments in which general image transmission is difficult described in Sec.~\ref{sec:background}. 
Fig.~\ref{fig:overview} shows an overview of the proposed method.
In the proposed method, to achieve image transmission with a small amount of transmitted data, the transmitter extracts and transmits multi-modal semantic features of the original image, and the receiver reconstructs an image with the same semantic information as the received features.
In the image generation process at the receiver, as the image-generation model generates various images, the receiver must select the best image as an output image.
To evaluate the reconstruction result, this study developed quality assessment procedures, which is introduced later.
In the proposed method, the receiver reconstructs images by the received semantic information to be easily recognized by humans, and therefore the proposed method can be potentially used in various applications.

To select an appropriate image as an output in the proposed method, the quality of the generated images must be assessed based on the semantic similarity between the original and generated images on the receiver.
Although some metrics assess the image quality by comparing two images~\cite{bib:lpips}, the proposed method cannot compare the original and generated images directly because the receiver does not possess the original image.
Therefore, the quality of the generated images must be assessed based only on the received semantic features of the original image.
To address this issue, this study proposes two scoring procedures for evaluating the similarity between semantic features, such as descriptive and segment information, of the original and generated images.
Additionally, this study also proposes a background recoloring method for enhancing the contours of objects in images to improve image generation.

The contributions of this study can be summarized as follows:
\begin{itemize}
    \item We propose an image construction method that reconstructs images at the receiver end using multi-modal semantic communication and similarity evaluations.
    Therefore, the proposed method can transmit images with higher semantic similarity information than conventional transmission methods.
    Additionally, in this paper, we employed only open-source machine learning algorithms~\cite{bib:blip,bib:deeplab,bib:stable-diffusion} for realizing the processes of the extraction of semantic features and image generation in the proposed method. This implementation highlights to facilitate the reproduction and adoption of the proposed method by others.
    \item Extending from~\cite{bib:ccnc}, this paper adds specific procedures to evaluate generated images by a receiver to select an output result among the images. The proposed method can select an image with high reproducibility as the output because the receiver evaluates the semantic similarity between the original and generated images to select the result.
    To realize this, we propose two scoring procedures for evaluating the semantic similarity between images.
    These procedures can evaluate the semantic similarity between images based only on the semantic features received by the receiver.
    \item We propose and investigate scoring procedures of the semantic similarity between the original and generated images. This is an important first step toward establishing evaluation metrics for image transmission methods using semantic communication.
    The results of the experiment indicate that the proposed procedures can evaluate semantic features of the target object, such as position and composition, and background information of the image.
    Additionally, the results demonstrate that the proposed method with the background recoloring technique can effectively reconstruct the composition and position of the target object in the generated images.
\end{itemize}

The remainder of this paper is organized as follows: 
Section \ref{sec:related} presents the existing image compression and transmission methods.
Section \ref{sec:proposed} introduces the proposed image-generation-based transmission methods.
Section \ref{sec:metrics} investigates evaluation metrics for evaluating semantic similarity between the original images and those generated using the proposed methods.
Section \ref{sec:experiment} elucidates the experiments performed to evaluate the proposed methods and the semantic similarity between images using the proposed metrics mentioned previous section.
Section \ref{sec:result} presents and discusses the experimental results.
Section \ref{sec:conclusion} concludes the paper and presents future research directions.

\section{Related Work}
\label{sec:related}

Many studies have investigated various machine-learning-based image compression methods~\cite{bib:learning-base-image-comp-1,bib:learning-base-image-comp-2,bib:cnn-base-image-comp-1,bib:cnn-base-image-comp-2,bib:cnn-base-image-comp-3,bib:ae-base-image-comp-1,bib:ae-base-image-comp-2}.
These methods compress the image data using various machine learning algorithms such as convolutional neural networks (CNNs)~\cite{bib:cnn-base-image-comp-1,bib:cnn-base-image-comp-2,bib:cnn-base-image-comp-3} and autoencoder~\cite{bib:ae-base-image-comp-1,bib:ae-base-image-comp-2}.
These methods have demonstrated higher compression ratios while maintaining better image quality than traditional image compression algorithms such as JPEG~\cite{bib:jpeg} and JPEG2000~\cite{bib:jpeg2000}.
However, these methods focus only on source coding to reduce the image data size.
When a device transmits images through a network, a channel coding algorithm must be used to prevent errors caused by channel noise. 

To address this issue, DeepJSCC-based methods were recently proposed to further improve image transmission efficiency~\cite{bib:deep-jscc-1,bib:deep-jscc-2}.
These methods encode an image in a batch using a CNN, which differs from the conventional procedures of employing separate source and channel coding steps.
These methods exhibit better performance than conventional image compression algorithms, such as JPEG and JPEG2000, for a specific channel bandwidth and a low signal-to-noise ratio.

The goal of these methods is to transmit visually complete images between end-to-end devices.
That is, these methods aim to transmit images while preserving all the semantic information visually recognizable by humans, including the composition, background, and target object type.
However, in some situations, it is redundant to transmit all semantic information contained in images.
Therefore, to significantly reduce the amount of data transmitted during image transmission, only some semantic information must be extracted from an image and transmitted to the receiver.
Recently, owing to the development of neural networks, various machine learning algorithms, such as image captioning~\cite{bib:blip,bib:clip} and semantic segmentation~\cite{bib:deeplab,bib:segnet}, that can extract specific semantic information from an image have been proposed to convert images into descriptive or segmented information.

This approach, which focuses on the transmission of semantic information, is known as semantic communication~\cite{bib:semantic-commun-1,bib:semantic-commun-2,bib:semantic-commun-3}.
Semantic communication aims to achieve end-to-end data transmission while maintaining the content at the semantic level rather than at the bit level.
In other words, semantic communication is considered to be successful when original and received data are semantically the same, even if the received data have been changed from the original data at the bit level.

\cite{bib:t2i-coding} proposed an image transmission method that transmits only the image caption to further improve communication efficiency.
In this method, a transmitter generates a caption as descriptive information of the original image using an image captioning algorithm.
Subsequently, the transmitter transmits only the generated caption, and the receiver reconstructs the image based on the received caption by using an image-generation model.
As this method extracts only the descriptive information from the original image, it may incur a loss of some semantic information contained in the original image that is difficult to represent through the descriptive information.
For example, it is difficult to represent the exact position and composition of a target object in an image using only descriptive information.
Therefore, to reconstruct an image more clearly at the receiver, the transmitter must extract and transmit multi-modal semantic information from the original image.

To address this issue, we previously proposed a concept of novel image transmission method that uses multi-modal semantic information~\cite{bib:ccnc}.
In this method, a transmitter extracts descriptive, segmented, and color information from the original image as semantic features, and transmits only these features.
After receiving these features, the receiver reconstructs the image using an image-generation model.
By conveying diverse information, such as the composition and color of the original image, in addition to descriptive information, this method enables image transmission that can generate images that are more semantically similar to the original image compared with a method using only single feature~\cite{bib:t2i-coding}. 
\cite{bib:ccnc} included only concept of the proposed method and subjective classification of the generated images as a preliminary study.
In the proposed method, the receiver needs to evaluate the generated images based on some metrics to select an output result among the images.
However, specific procedures and investigating evaluation metrics for these evaluation and result selection processes are not included in~\cite{bib:ccnc}.

To evaluate the overall performance of image-compression algorithms, the peak signal-to-noise ratio (PSNR) is often employed as a metric to assess the visual similarity between two images. 
PSNR is defined as the ratio of the maximum power of a signal to the noise that degrades the image quality.
Another metric, called the learned perceptual image patch similarity (LPIPS)~\cite{bib:lpips}, has been employed in existing semantic communication methods for image transmission~\cite{bib:t2i-coding}.
LPIPS focuses on the distance in the latent space and compares the feature output of the convolution layers of a trained neural network model for image classification. 

These metrics are useful for evaluating methods that aim to fully reconstruct the original image.
However, the proposed image transmission method aims to reconstruct an image using specific semantic features required by the receiver and extracted from the original image.
In other words, the proposed method does not involve transmitting visually complete images identical to the original image.
Therefore, it is necessary to evaluate the semantic similarities between the original image and the images generated using the proposed method to compare them.
However, evaluating semantic similarity using the conventional evaluation metrics is challenging. 

\section{Image Generation-based Transmission Method}
\label{sec:proposed}

\begin{figure}[b]
  \centering
  \includegraphics[clip,width=0.99\linewidth]{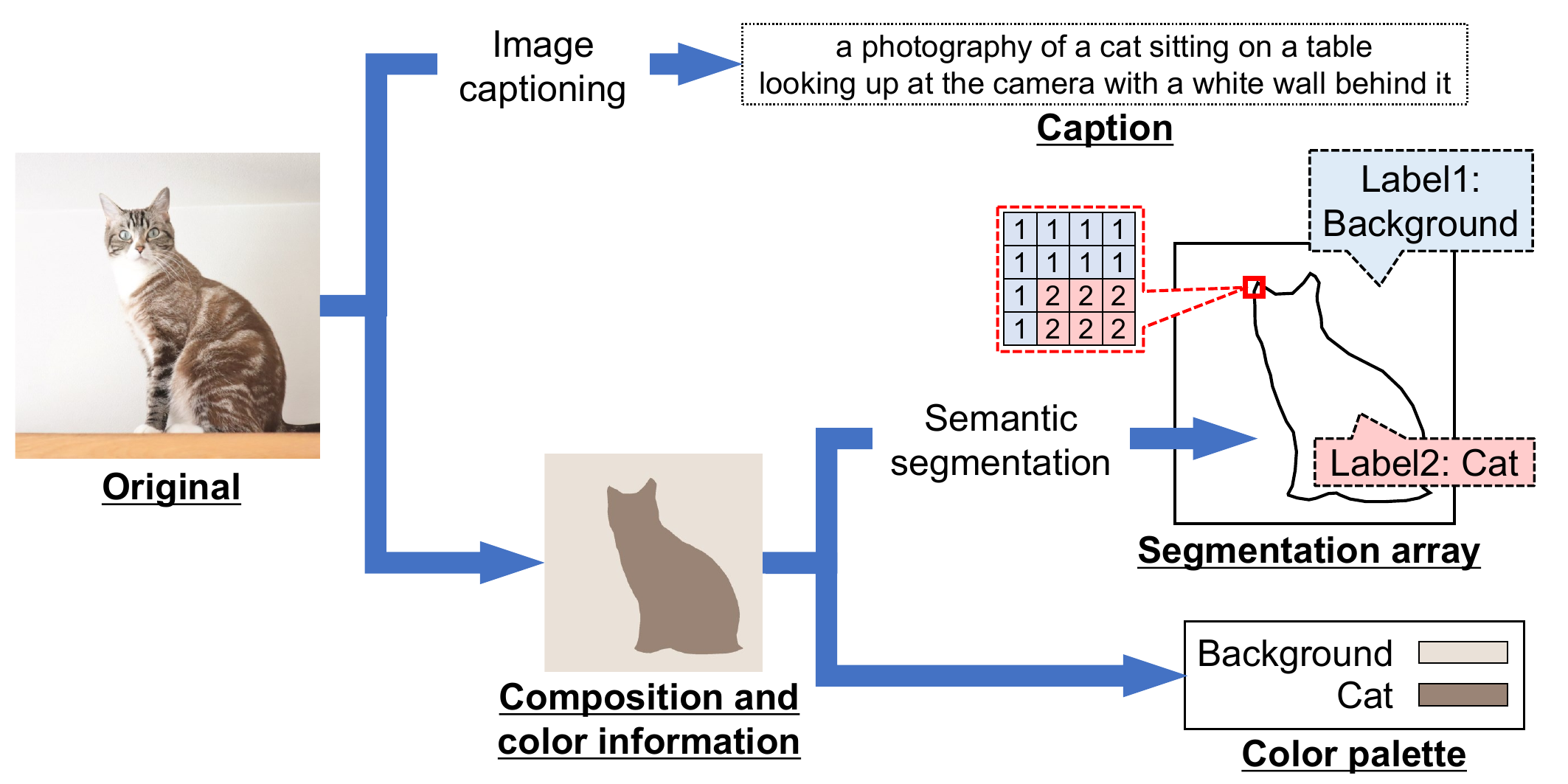}
  \caption{Procedure for extracting semantic information from the original image using the transmitter.
  The transmitter generates a caption, segmentation array, and color palette from the original image, representing its descriptive, segment, and color information.}
  \label{fig:sender}
\end{figure}

\begin{figure}[t]
  \vspace{3mm}
  \centering
  \includegraphics[clip,width=0.99\linewidth]{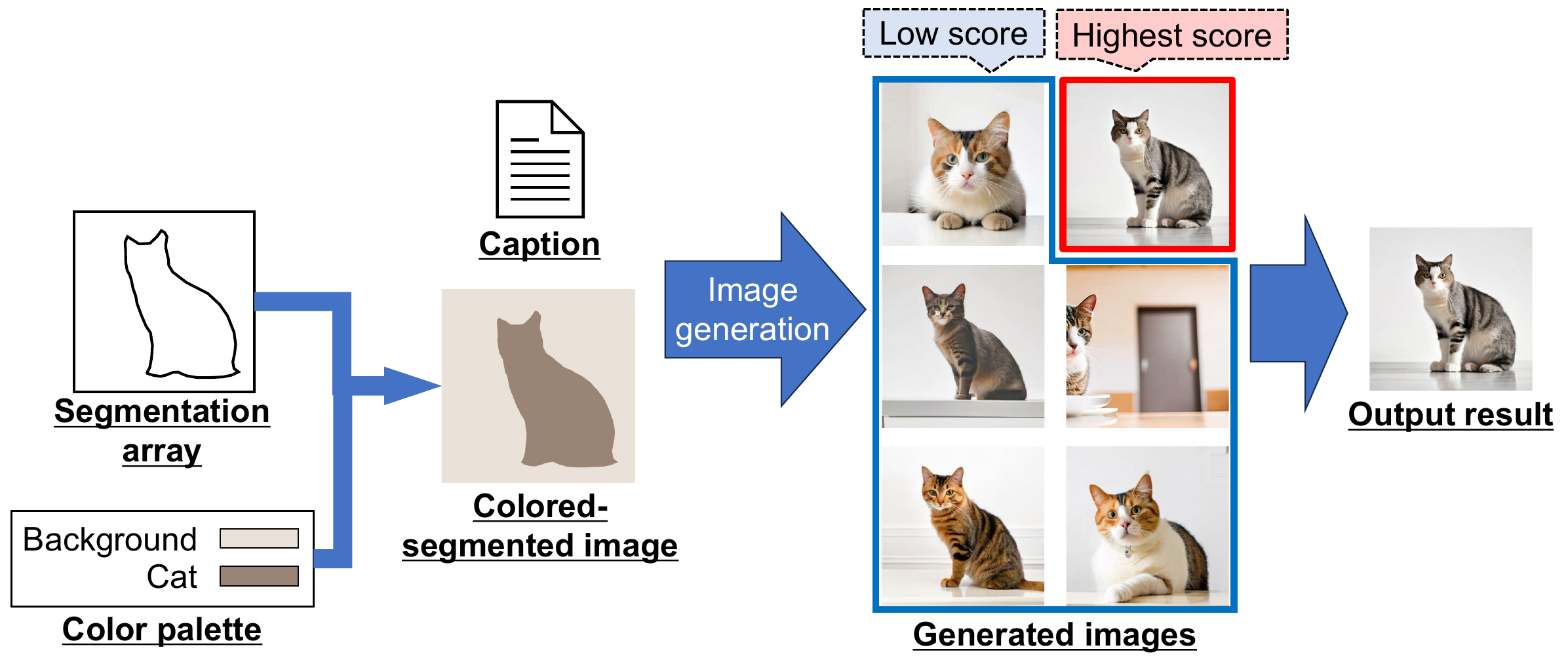}
  \caption{Procedure for reconstructing the image at the receiver using the image-generation model.
  The receiver generates multiple  images by inputting the received caption and generated segmented-colored image into an image-generation model.
  After the generation, the receiver selects an output image based on the semantic similarity between the received features and those of the generated images.}
  \label{fig:receiver}
\end{figure}

This section describes the proposed image generation-based transmission method that uses multi-modal semantic information.
A transmitter extracts multiple semantic features, such as composition, description, and color information, representing diverse semantic information from an image and transmits only these extracted features.
The details of these extracted features are explained in Sec.~\ref{sec:transmitter}.
The receiver generates multiple images using the received features and selects an output image based on the semantic similarity.
The details of these procedures at the receiver are explained in Sec.~\ref{sec:receiver}.
The proposed method focuses on both a reduction of transmitted data size and image reconstruction while maintaining specific semantic features required by receiver's application.

\subsection{Feature Extraction at the Transmitter}
\label{sec:transmitter}

Fig.~\ref{fig:sender} illustrates the process employed by the transmitter for extracting multi-modal semantic features from the original image (size $m \times n$ pixels) using existing machine learning algorithms.

First, the transmitter generates a caption as the descriptive information using an image captioning algorithm~\cite{bib:coca,bib:clip,bib:blip}.
The descriptive information contains the target object type and broad background information of the original image.

Next, the transmitter performs semantic segmentation~\cite{bib:deeplab,bib:segnet} on the original image to obtain a segmentation array that represents the segment information of the image.
Semantic segmentation algorithms realize image segmentation by labeling each pixel of an image.
Therefore, the segmentation array is an $m \times n$ two-dimensional array composed of labels for all pixels of the original image.
The segment information represents the position and composition of the target object in the original image.

Third, the transmitter generates a color palette by calculating the average RGB value for each label in the segmentation array.
This is because the detailed color information of the image is lost when only the caption and segmentation array are generated, and it is difficult to reconstruct the color information of the image by the receiver.
The color palette represents the color information of the segments contained in the original image.
After extracting the three types of semantic features, the transmitter transmits this data, the size of which is significantly smaller than that of the original image, to the receiver.

\subsection{Image Reconstruction by the Receiver}
\label{sec:receiver}

Fig.~\ref{fig:receiver} illustrates the image reconstruction procedure employed by the receiver.
After receiving the semantic features, the receiver reconstructs the image using the received information and an image-generation model.
First, the receiver generates a colored-segmented image from the segmentation array and color palette, which is then input into the image-generation model.
The colored-segmented image is created by changing the value of each label in the segmentation array to its corresponding RGB value in the color palette.
Subsequently, the receiver inputs the received caption and generated colored-segmented image into the image-generation model to generate multiple images.
Here, the receiver must evaluate the semantic similarities between the original and generated images to output the best result among the generated images.
However, it is not possible to directly compare the similarities between the images because the receiver does not have the original image.
To address this issue, the receiver assigns similarity scores based on the received semantic features and generated images.
The detailed scoring procedures for semantic similarity are discussed below.
After scoring, the receiver selects the image with the highest score as the output.
Thus, the proposed method realizes image transmission based on the semantic information contained in an image using these procedures.

\begin{figure}[b]
  \centering
  \includegraphics[clip,width=0.9\linewidth]{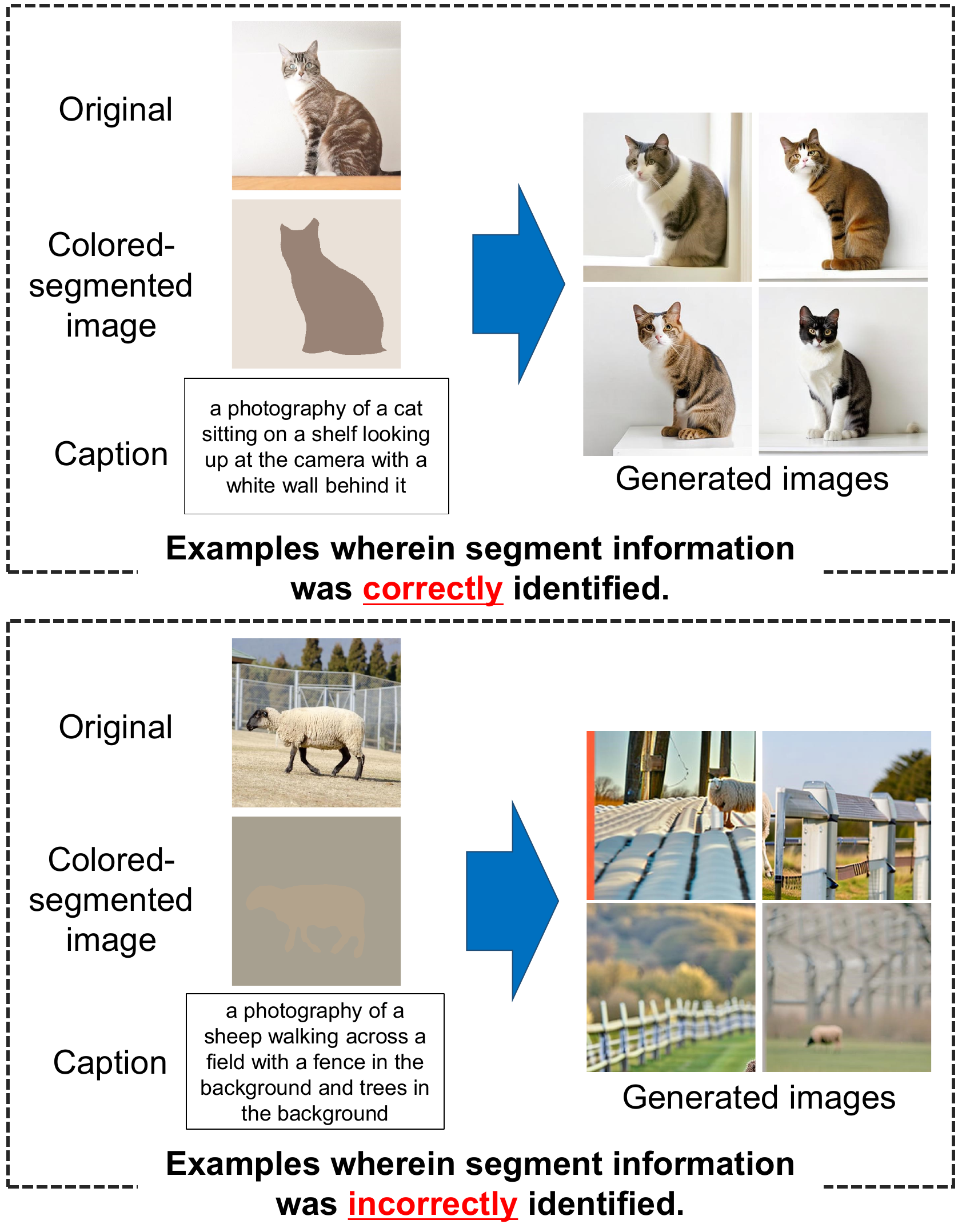}
  \caption{Examples wherein color information affects the segment information recognition of the image-generation model.
  In the upper example, the image-generation model correctly recognizes the composition of the target object because the boundary between the object and background in the colored-segmented image is clear.
  In contrast, in the lower example, the positions of the target object in the generated images differ significantly from that in the original image, which may be caused by the similar RGB values of the target object and background in the colored-segmented image.}
  \label{fig:seg-example}
\end{figure}

\subsection{Background Recoloring for Enhancing Object Contours}

As shown in Fig.~\ref{fig:seg-example}, the image-generation model cannot accurately identify the segment information if similar RGB values are assigned to the ``background'' and target object labels. 
Therefore, the receiver generates images with completely different compositions than that of the original image when using this segment information.
To avoid this undesirable image generation, this study proposes an extensive approach that changes the color of pixels labeled ``background'' to white.
When employing this approach, the segment information of the original image is more likely to be recognized by the image-generation model because the area of the target object in the colored-segmented image stands out. 
However, the background color information of the original image may be lost because the background color of the colored-segmented image is converted to white.

\section{Evaluation Metrics for the Proposed Method}
\label{sec:metrics}

This section describes the similarity scoring procedures for selecting an output from the generated images. 
As described in Sec.~\ref{sec:receiver}, to score the generated images, the receiver requires similarity metrics.
However, the receiver cannot directly compare the similarity of the images because it cannot access the original image.
To address this issue, this study proposes two scoring procedures to compare the semantic similarity between the received features of the original image and those of the images generated by the receiver.

\textbf{Scoring procedure for descriptive information similarity:} This procedure compares the descriptive information in the original images and those generated by the receiver.
The receiver generates captions for the generated images and calculates the text similarity between them and that of the original image.
This procedure uses BERTScore~\cite{bib:bert-score}, which is a metric used to assess the semantic similarity between sentences to evaluate text similarity.
BERTScore calculates text similarity using bidirectional encoder representations from transformers (BERT)~\cite{bib:bert}, which is a natural language processing model.
Text similarity is calculated based on the vectors of input sentences generated by the trained BERT.

\textbf{Scoring procedure for segment similarity:} This procedure compares the segment information contained in the original and generated images.
The receiver first converts the generated images into segmentation arrays using a semantic segmentation algorithm, similar to the transmitter.
As described in Sec.~\ref{sec:proposed}, each element of the segmentation array contains a label for the corresponding pixel.
Note that the segmentation arrays of the original image $S_{\textrm{org}}$ and the generated image $S_\textrm{seg}$ are defined as follows:
\begin{equation}
    \label{eq:array}
    \begin{split}
        S_{\textrm{org}} \coloneq (p_1, p_2, \cdots p_N) \\
        S_{\textrm{seg}} \coloneq (q_1, q_2, \cdots q_N),
    \end{split}
\end{equation}
where $p$ and $q$ denote the labels stored in each element of the segmentation arrays, that is the corresponding pixels.
Additionally, $N \coloneq mn$ denotes the size of segmentation arrays, i.e. the number of pixels of the original and generated images, which contain, the same number of pixels.
Subsequently, the receiver calculates the segmentation matching rate $\textrm{SMR}$ between $S_{\textrm{org}}$ and $S_{\textrm{seg}}$ as follows: 
\begin{equation}
    \label{eq:smr}
    \textrm{SMR} \coloneq \frac{1}{N} \sum_{i=1}^N \delta(p_i, q_i).
\end{equation}
where $\delta(p_i, q_i)$ represents the Kronecker delta that indicates whether the two arguments are the same, and is defined as follows:
\begin{equation}
    \label{eq:match}
    \delta(x, y) \coloneq 
    \begin{cases}
    1 & (x = y) \\
    0 & (x \neq y).
    \end{cases}
\end{equation}
The receiver calculates $\textrm{SMR}$ by determining the percentage of pixels with matching labels among all pixels.
$\textrm{SMR}$ can compare the position and composition of the target object between the images.

\begin{figure}[b]
  \vspace{3mm}
  \centering
  \includegraphics[clip,width=0.99\linewidth]{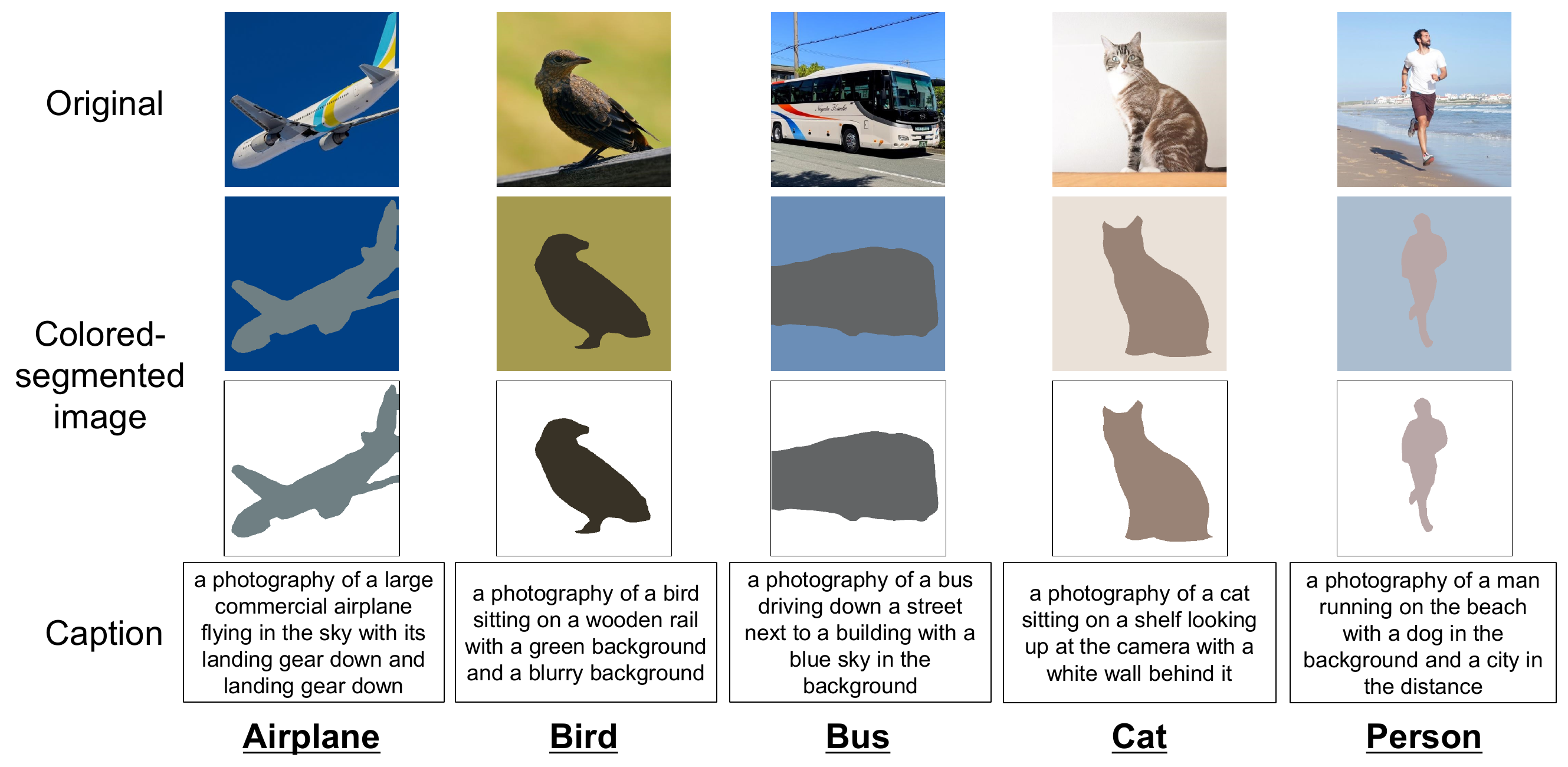}
  \caption{Examples of original images, segmented colored images, and captions used in the experiment.}
  \label{fig:seg-text-example}
\end{figure}

\begin{table*}[t]
    \renewcommand{\arraystretch}{1.3}
    \centering
    \caption{Data sizes employed by each method. 
    The average data size of the captions is approximately $1/270$ of that of the images in the JPEG format.
    Additionally, even when multiple semantic features are combined, the data size is approximately $1/14$ of that of the JPEG format images.}

    \begin{tabular}{cccc}
        \Hline
        Uncompressed & JPEG & Caption & 
        \begin{tabular}{c}
             Caption + color palette\\ + segmentation array
        \end{tabular}\\ \hline
        786 [KB] & 41.8 [KB] & 0.0998 [KB] & 2.90 [KB]\\
        \Hline
    \end{tabular}
    \label{tab:data-size}
\end{table*}

\section{Experiment Setup}
\label{sec:experiment}

We conducted an experiment to evaluate the performances of the proposed methods and scoring procedures for determining the similarity between the semantic features of the original and generated images.
This experiment included 105 images; five each containing 21 different objects (airplane, bicycle, bird, boat, bottle, bus, car, cat, chair, table and chairs, cow, dog, horse, motorbike, person, potted plant, sheep, sofa, table, train, and TV).
All images were in the JPEG format with a quality of 80 and size of $512 \times 512$ pixels, and were obtained from photoAC~\cite{bib:photoAC}, which contains copyright-free images.

First, all the images were converted into captions and segmentation arrays using image captioning and semantic segmentation algorithms.
This experiment used bootstrapping language-image pre-training for unified vision-language understanding and generation (BLIP)~\cite{bib:blip} for image captioning and DeepLabV3~\cite{bib:deeplab} for semantic segmentation.
In addition, color palettes for each image were simultaneously created based on the segmentation arrays.
Note that BLIP was set to generate captions containing 20--30 words beginning with ``a photography of'' because the experiment employed only photos.
After conversion, the colored-segmented images were created in the PNG format using the segmentation arrays and corresponding color palettes.
Fig.~\ref{fig:seg-text-example} shows examples of the colored-segmented images and captions used in this experiment.
Subsequently, 50 images were generated for each original image by inputting the colored-segmented image and the caption into Stable Diffusion~\cite{bib:stable-diffusion}, which is a widely used image-generation model.
Note that in this experiment, Stable Diffusion was instructed to avoid the following factors to prevent generating images with textures different from the original images.

\begin{figure*}[t]
\vspace{-5mm}
  \centering
  \begin{tabular}{c}
    \begin{minipage}{0.32\hsize}
      \centering
      \includegraphics[clip, width=0.9\textwidth]{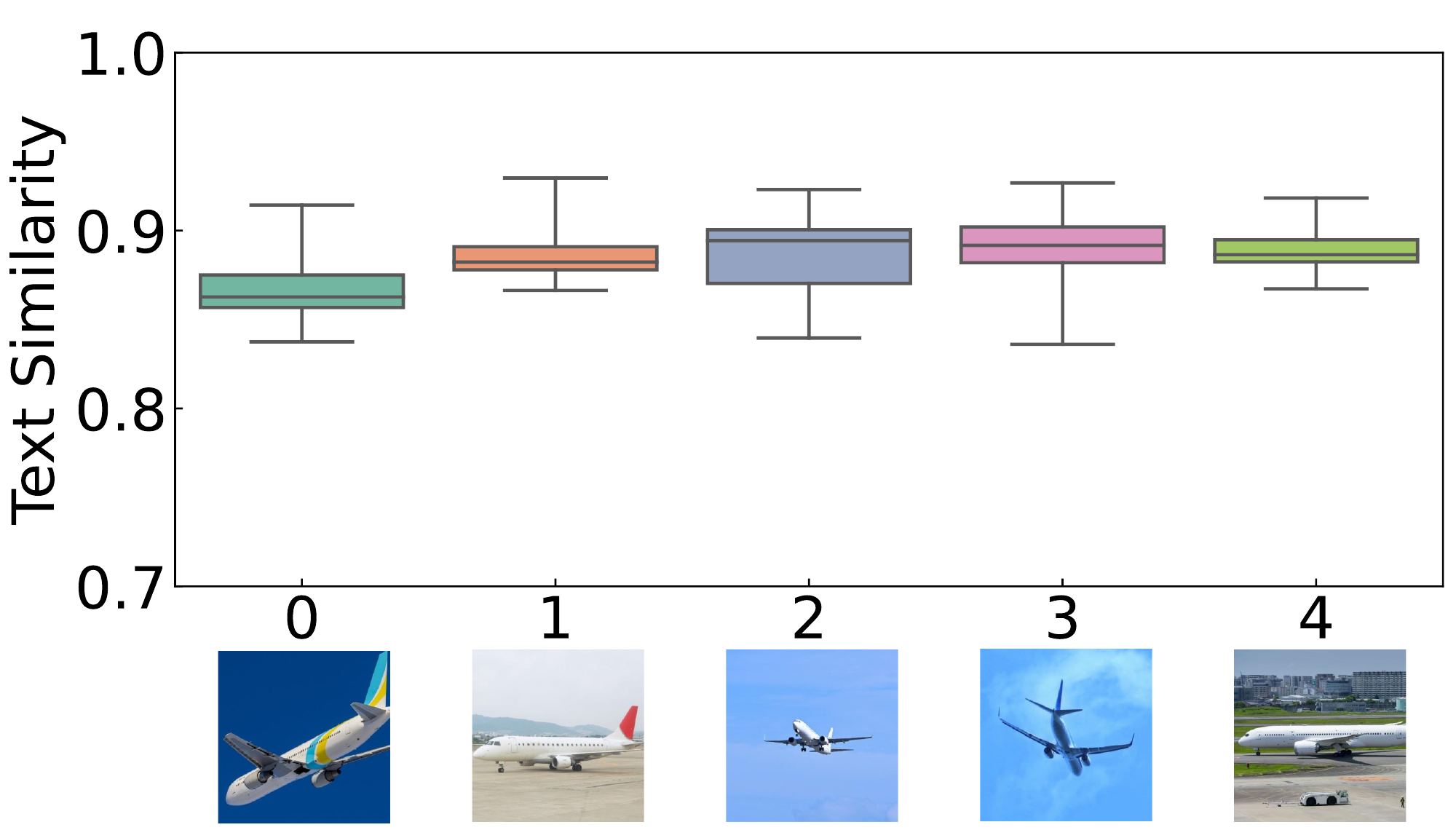}
    {\footnotesize
      \\ (a)~Using only the caption.}
    \end{minipage}
    \begin{minipage}{0.32\hsize}
      \centering
      \includegraphics[clip, width=0.9\textwidth]{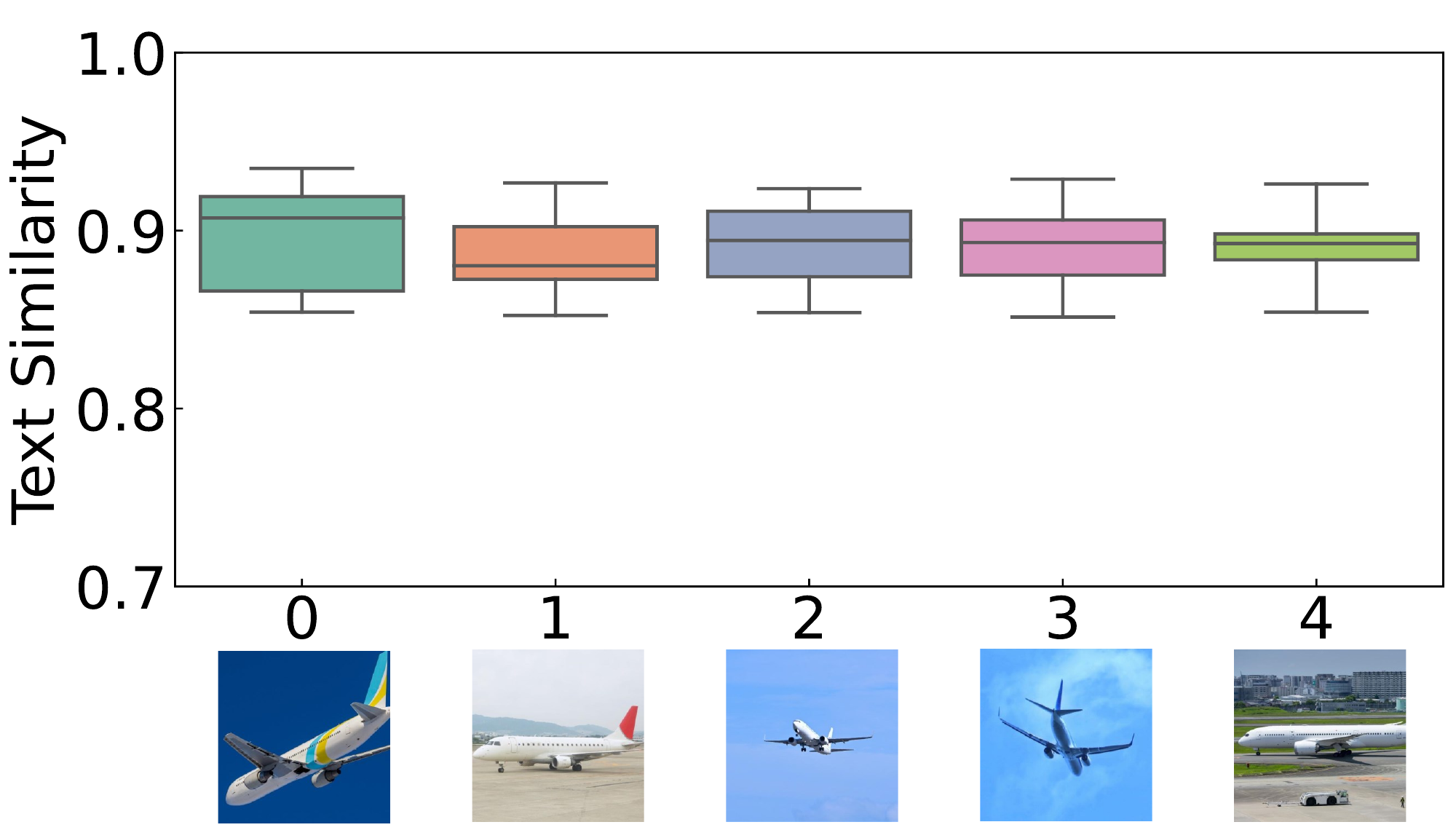}
    {\footnotesize
      \\ (b)~Proposed method w/o background recoloring.}
    \end{minipage}
    \begin{minipage}{0.32\hsize}
      \centering
      \includegraphics[clip, width=0.9\textwidth]{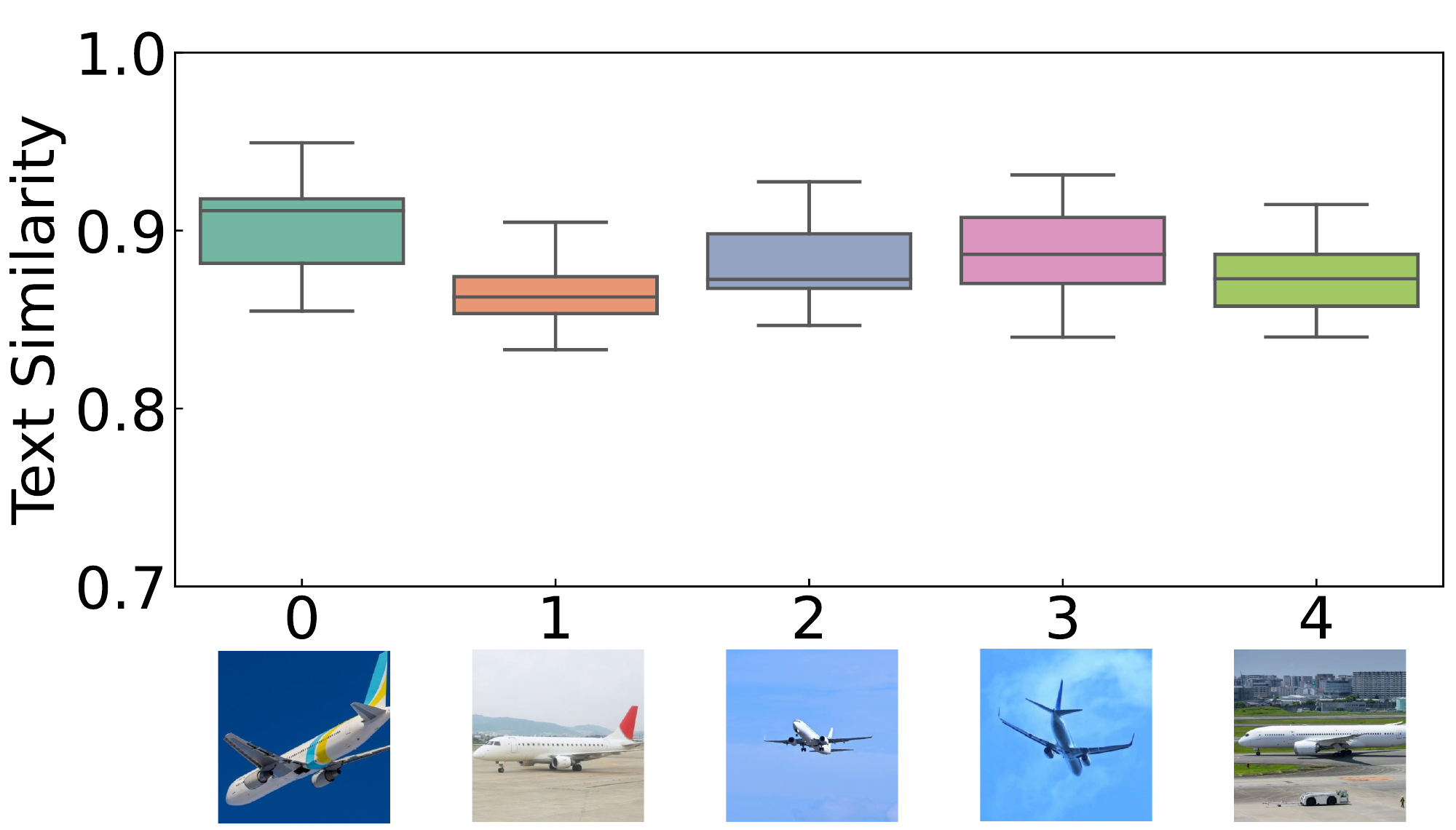}
    {\footnotesize
      \\ (c)~Proposed method with background recoloring.}
    \end{minipage}
  \end{tabular}
  \vspace{2mm}
  \caption{Text similarity scores of the ``airplane'' images generated by each method. 
  Note that the stop words are not removed in these results.
  The original images are shown at the bottom of the graphs.
  The results of the proposed method with background recoloring exhibit lower similarity scores than those obtained using the method without background recoloring.
  This is because the background color information in the original image is lost via background recoloring.}
  \label{fig:bert-score-1}
  \vspace{-3mm}
\end{figure*}

\begin{figure}[t]
  \centering
  \includegraphics[clip,width=0.75\linewidth]{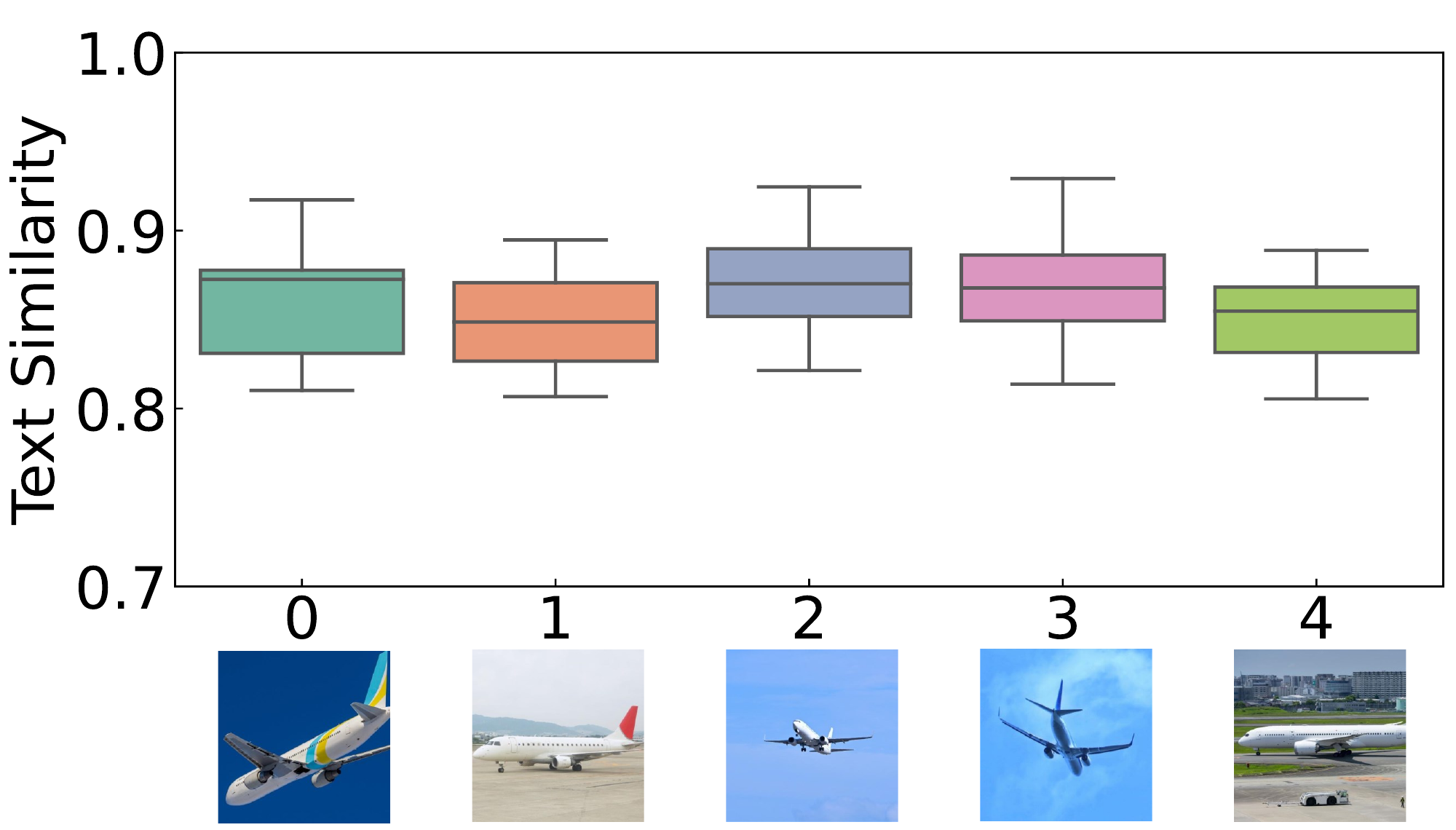}
  \caption{Text similarity scores of the ``airplane'' images generated by the proposed method without background recoloring.
  Note that the stop words are removed in the results.
  Compared with the results of not removing the stop words, these results exhibit lower similarity scores and larger variances.
  This is because removing the stop words prevents the text similarity scores from becoming excessively high.}
  \label{fig:bert-score-2}
\vspace{-3mm}
\end{figure}

\noindent
\textbf{``low quality, worst quality, out of focus, ugly, error, jpeg artifacts, lowers, blurry, broken, illustration, animation, painting, 2D, oil painting, sketch, watercolor, ink, flat color''}

We compared the semantic features of the original images and those generated by each method using the scoring procedures described in Sec.~\ref{sec:metrics}.
Note that in the text similarity evaluation, we compared the performance with and without removing stop words from the captions.
Stop words are prepositions and articles that are removed during preprocessing for natural language processing as their presence can result in excessively high similarity scores.
We also evaluated and compared the performance of the simplest method, which generated images using only the caption generated from the original images, with that of the two methods proposed in Sec.~\ref{sec:proposed}.

\begin{figure*}[t]
  \centering
  \begin{tabular}{c}
    \begin{minipage}{0.32\hsize}
      \centering
      \includegraphics[clip, width=0.9\textwidth]{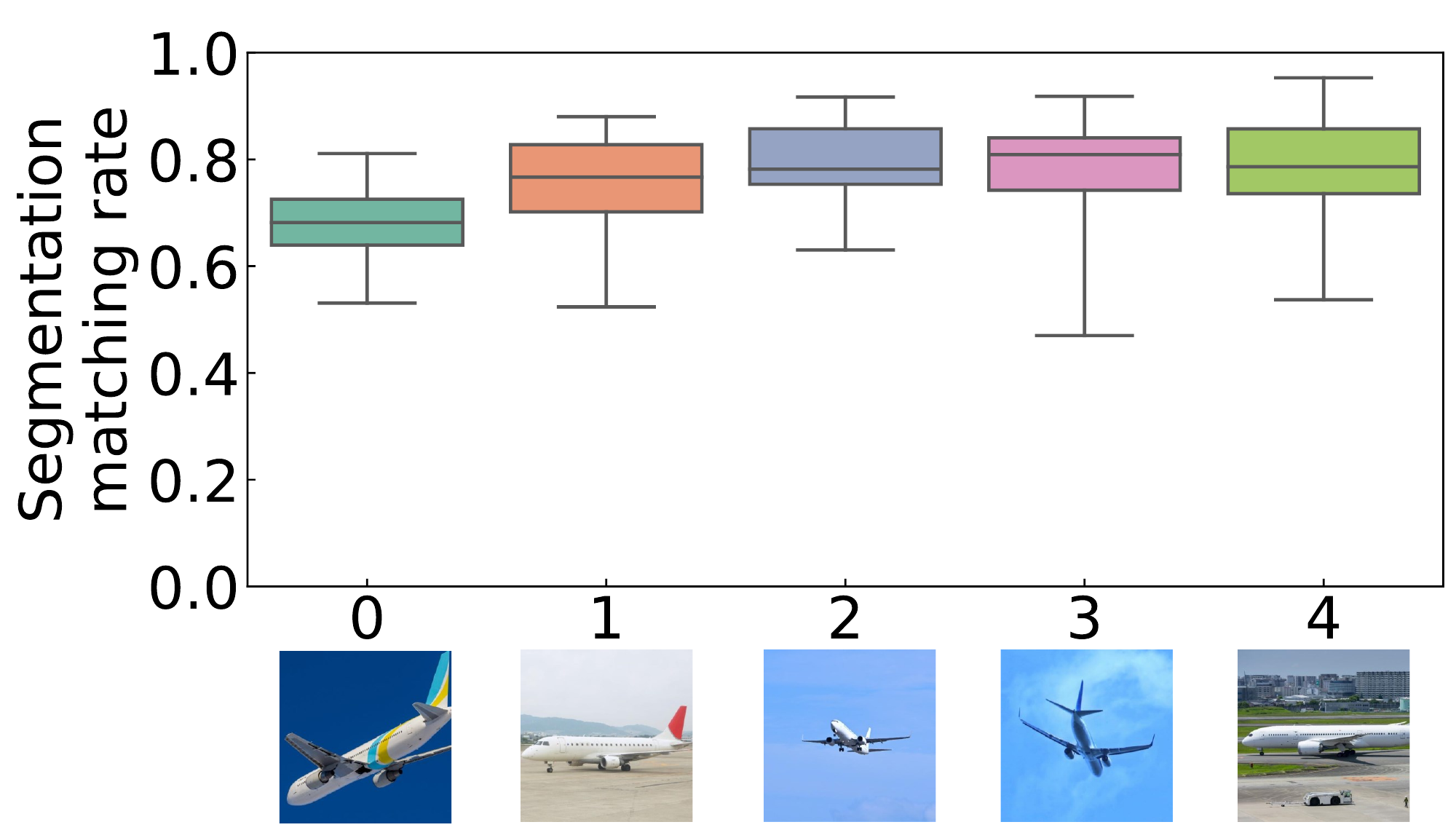}
    {\footnotesize
      \\ (a)~Using only captions.}
    \end{minipage}
    \begin{minipage}{0.32\hsize}
      \centering
      \includegraphics[clip, width=0.9\textwidth]{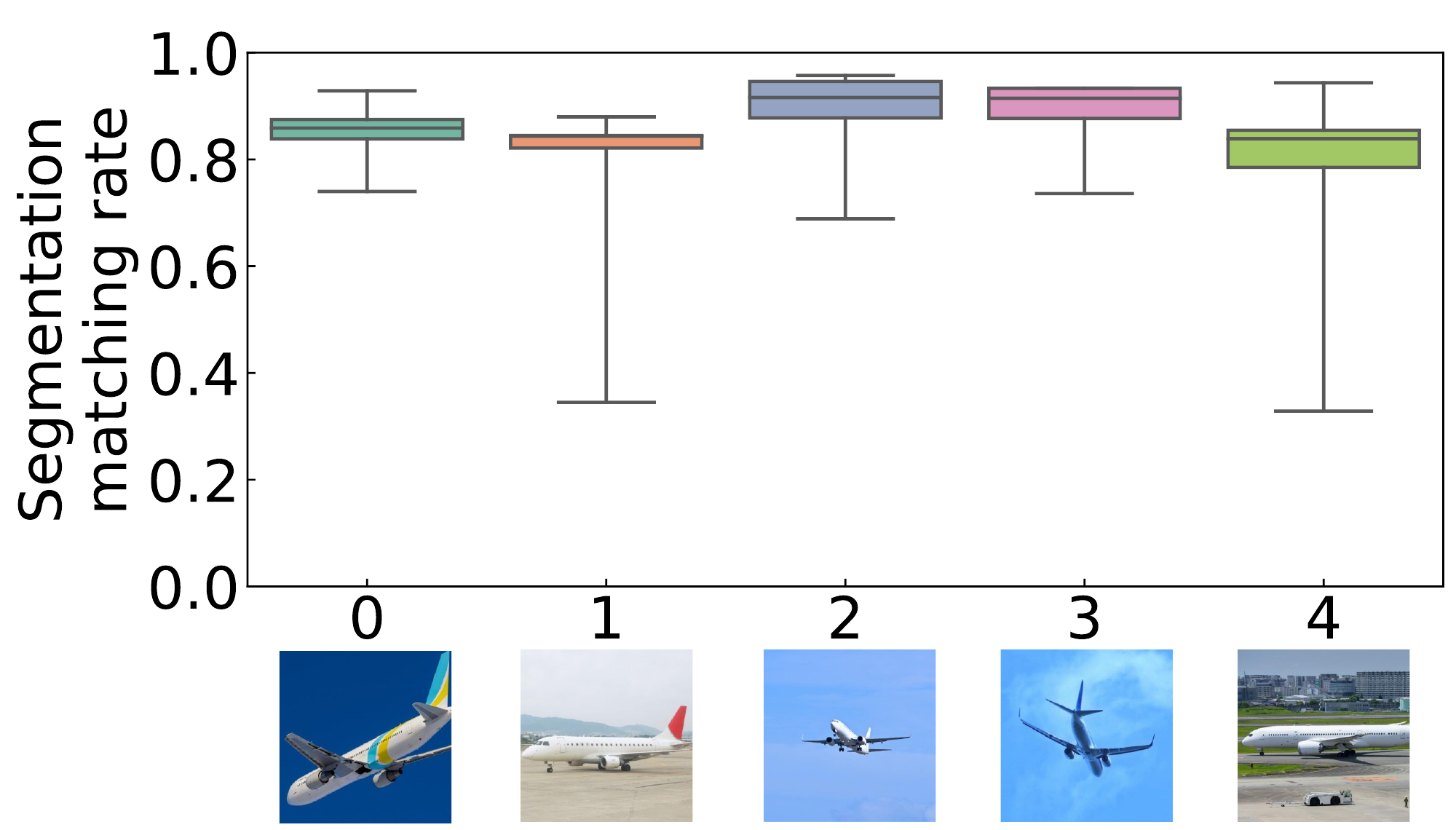}
    {\footnotesize
      \\ (b)~Proposed method w/o background recoloring.}
    \end{minipage}
    \begin{minipage}{0.32\hsize}
      \centering
      \includegraphics[clip, width=0.9\textwidth]{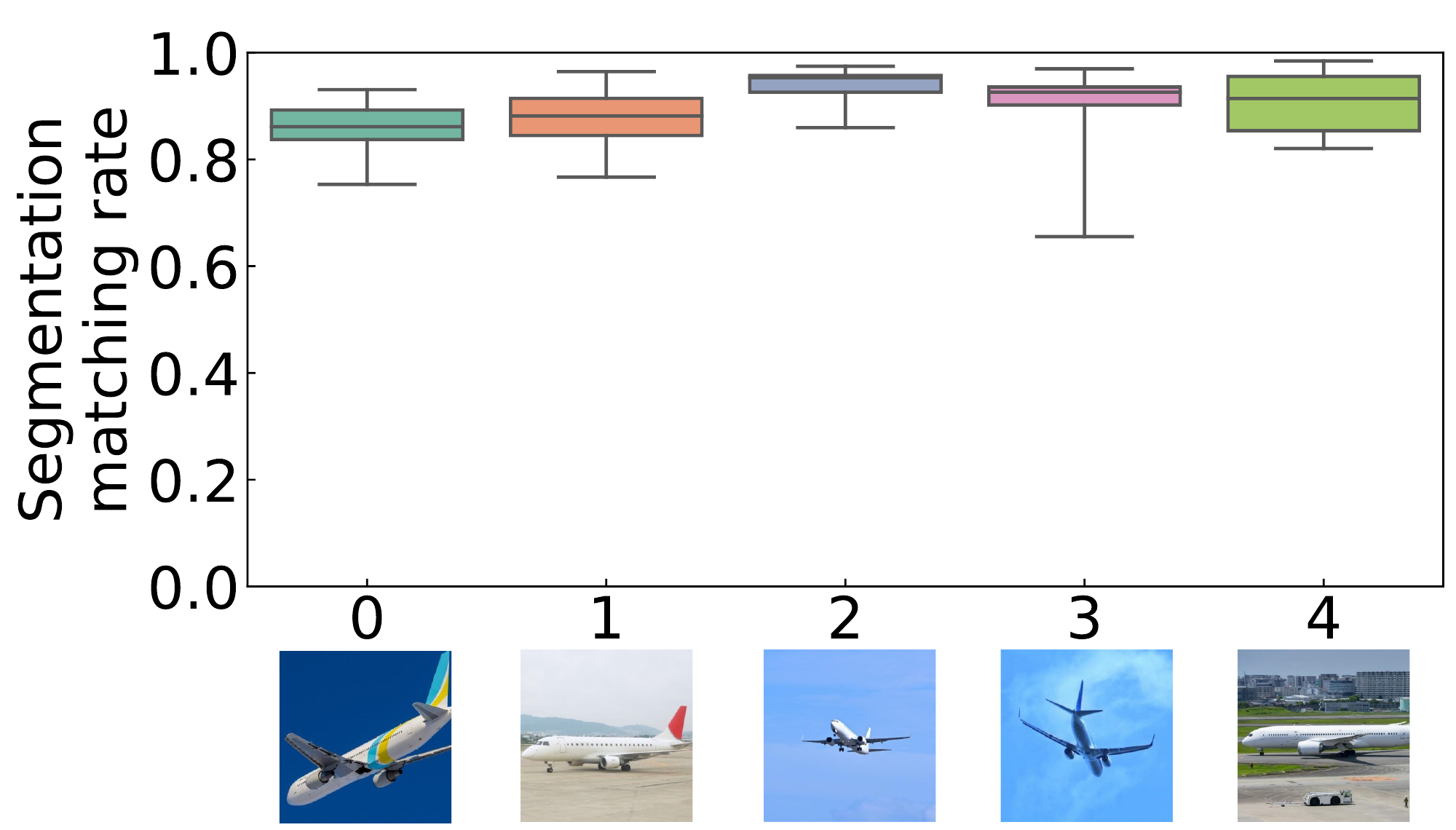}
    {\footnotesize
      \\ (c)~Proposed method with background recoloring.}
    \end{minipage}
  \end{tabular}
  \vspace{2mm}
  \caption{Segmentation matching rates of the ``airplane'' images generated using each method.
  The original images are shown at the bottom of the graphs.
  The results of the proposed methods exhibit higher matching rates than those obtained using only captions.
  Additionally, the proposed method with background recoloring exhibits a higher matching rate than that without background recoloring.
  This is because background recoloring increases the segment information recognition performance of the image-generation model.}
  \label{fig:seg-match-1}
\vspace{-3mm}
\end{figure*}

\begin{figure}[t]
  \centering
  \includegraphics[clip,width=0.99\linewidth]{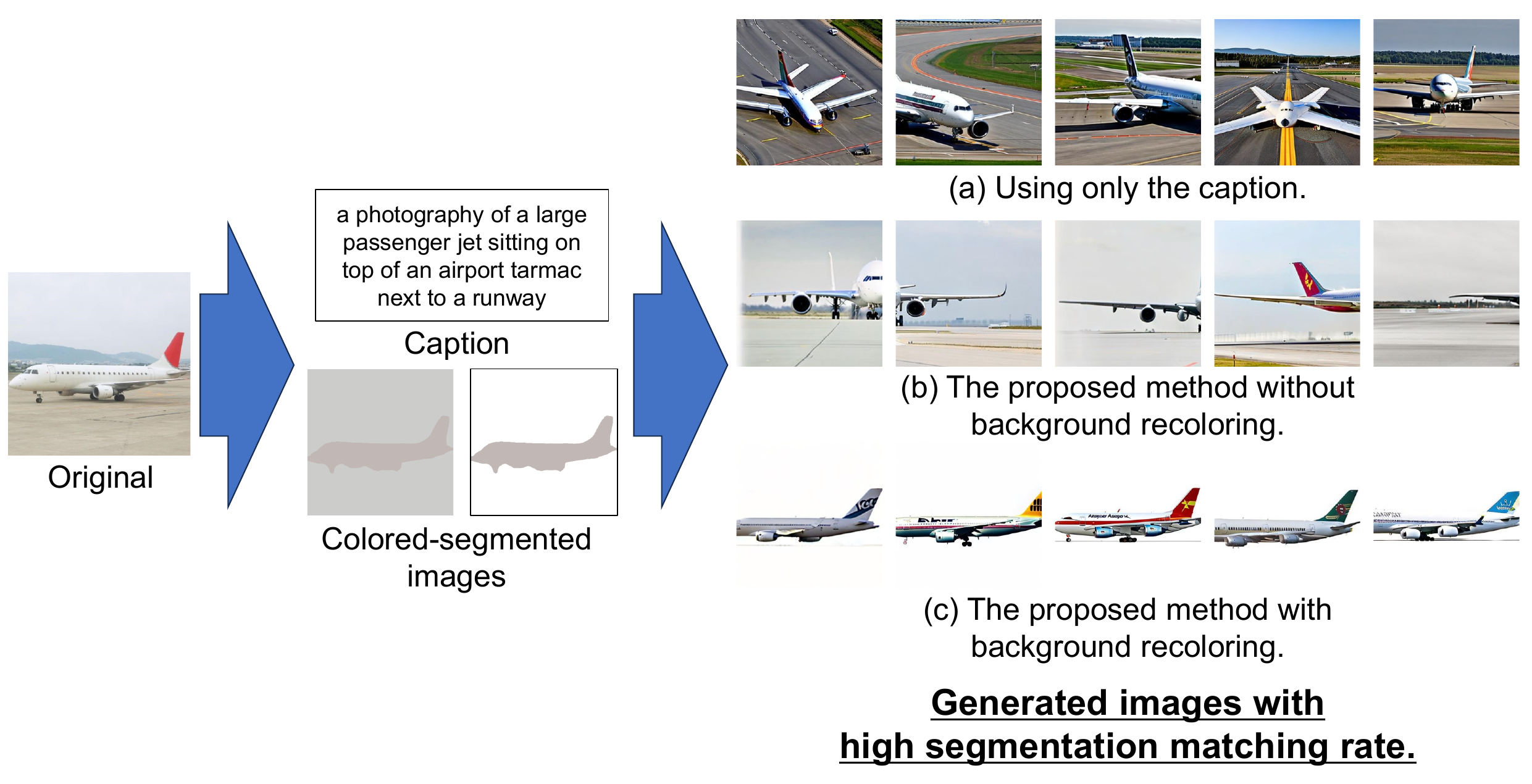}
  \caption{Top five generated images with the highest segmentation matching rates with the original ``airplane'' image.
  Compared to the images generated using the method with background recoloring, the compositions of the images generated using the proposed method without background recoloring are less similar to that of the original image.
  The image-generation model cannot recognize the segment information using the proposed method without background recoloring owing to the similar RGB values of the target object and background.
  In such situations, background recoloring can generate images with more similar compositions to the target object.}
  \label{fig:match-rate-rank-airplane-1}
\end{figure}

\section{Experiment Results and Discussion}
\label{sec:result}

This section presents and discusses the results of the experiment described in Sec.~\ref{sec:experiment}, wherein the following aspects.
First, the reduction in transmission data size using the proposed method was evaluated in Sec.~\ref{sec:data-size}.
Second, the image-generation performances of each method were compared using the proposed scoring procedures in Sec.~\ref{sec:text-similarity} and Sec.~\ref{sec:seg-match}.
Third, the impacts of the content and combinations of semantic features on the generated images are discussed in Sec.~\ref{sec:characteristirc} and Sec.~\ref{sec:caption}.
Finally, the advantages and effectiveness of the proposed method and scoring procedures are discussed in Sec.~\ref{sec:scoring-advantage} and Sec.~\ref{sec:effectiveness}.

\subsection{Data Sizes of Semantic Features}
\label{sec:data-size}

We compared the data size transmitted by the proposed method with that of the raw bit maps, JPEG formatted data, and generated captions of the original images to validate its communication efficiency.
The reason we focused on transmission data size is that the proposed transmission method aims to realize a reduction of transmitted data size, and the reduction in transmission data size will enhance network performances such as shorter transmission delay and reduced network congestion. 
Therefore, this evaluation result indirectly confirms that effectiveness of the proposed method in environments where network resources are limited.

Table~\ref{tab:data-size} presents the data sizes for each compression approach.
``Uncompressed'' denotes the size of the uncompressed bit map of the original images, ``JPEG'' denotes the average size of the original images in the JPEG format, and ``Caption'' denotes average size of the captions for each original image, which is used by the simple method employing only the descriptive information of the images.
In addition, ``Caption + color palette + segmentation array'' denotes the average total size of the caption, color palette, and segmentation array for each image, which is the size of semantic features used in the proposed method.
The sizes of all elements of the segmentation arrays and captions characters were calculated as 1 byte each.
The data size of each color palette was calculated as the number of labels in an image $\times$ 4 bytes: 3 bytes for the RGB values and 1 byte for the label index.
Each segmented array was compressed using run-length encoding for further data reduction.
Additionally, ``Caption + color palette + segmentation array'' includes the RGB value of the ``background'' label.

The results indicate that the average data size of the captions is approximately $1/270$ of that of the images in the JPEG format.
Additionally, even when multiple semantic features are combined, the data size is approximately $1/14$ of that of the JPEG format images.
This result indicates that the proposed method can achieve higher efficiency in image transmission compared to general image compression algorithms.
For example, LoRaWAN that is one of LPWA technologies provides data rate of up to 37.5 kbps in LoRa modulation~\cite{bib:lpwa}.
This means that it takes approximately 8.9 seconds to transmit one JPEG image at this data rate.
In contrast, it takes about 0.62 seconds to transmit semantic features equivalent to one image with the proposed method, thereby underscoring the utility of the proposed method in situations with limited network resources.
Additionally, data size comparisons of the semantic features showed that only the caption extracted from the images had a smaller data size than the total size of the multiple semantic features, indicating that the communication efficiency was better when only captions were used.
However, the proposed method was superior in terms of color and composition reproduction, as described in the following sections.

\subsection{Comparison of Text Similarity Scores of Each Method}
\label{sec:text-similarity}

Fig.~\ref{fig:bert-score-1} shows the text similarity scores for the captions generated from the original image and 50 images generated using each method.
The results indicate the text similarity scores of the five original images of ``airplane,'' and similar trends were observed for the results of other images.
Note that stop words were not removed from these results and the original images are shown at the bottom of the figures.
The results demonstrate that text similarity was higher when multiple semantic features were used for image generation than when only captions were used for some original images.
This is because the proposed method generates images with more similar semantic information, such as the composition of the object and color information, using multiple semantic features.
In addition, the method that considers segment information (i.e., when the background color of the colored-segmented image is white), exhibited lower text similarity for some images compared to the method that does not consider segment information.
This is because, in the method that considers segment information, the color information of the background in the original image is lost, which may affect its text similarity performance.

Fig.~\ref{fig:bert-score-2} shows the text similarity scores for the captions generated from the original image and 50 images generated using the proposed method without background recoloring.
Note that the stop words were removed from all the captions in this result.
Therefore, this result can be compared to that shown in Fig.~\ref{fig:bert-score-1} (b) based on the presence or absence of stop words.
Fig.~\ref{fig:bert-score-1} (b) shows high text similarity scores for all original images, which may be caused by the presence of stop words that make similarity excessively high. 
In contrast, in Fig.~\ref{fig:bert-score-2}, the similarity scores for all images are lower than that in Fig.~\ref{fig:bert-score-1}; however, their variances are higher because of the removal of stop words.
Therefore, these results suggest that removing stop words is an effective strategy to improve the evaluation accuracy of the descriptive information in images.

\subsection{Comparison of Segmentation Matching Rate of Each Method}
\label{sec:seg-match}

Fig.~\ref{fig:seg-match-1} shows the segmentation matching rates for the segmentation arrays generated from the original five ``airplane'' images and the 50 images generated using each method.
Approximately, the same trends can be observed in the results of the other images, wherein the proposed methods exhibit higher matching rates than using only captions for the image generation.
This is because the proposed methods can generate images with similar compositions and positions of the target object as those in the original image using the colored-segmented images.

Among the proposed methods, that without background recoloring exhibits a large variance in the matching rate for some images, whereas that with background recoloring maintains a high matching rate for all images.
This is because, as shown in Fig.~\ref{fig:match-rate-rank-airplane-1} (b), when the colors of the background and target objects are similar, the image-generation model cannot correctly recognize the object position and composition.
By contrast, when the background color was set to white, the generated images had a composition that was more similar to the original image than that generated using the method without background recoloring as shown in Fig.~\ref{fig:match-rate-rank-airplane-1} (c).

\begin{figure}[t]
  \centering
  \includegraphics[clip,width=0.99\linewidth]{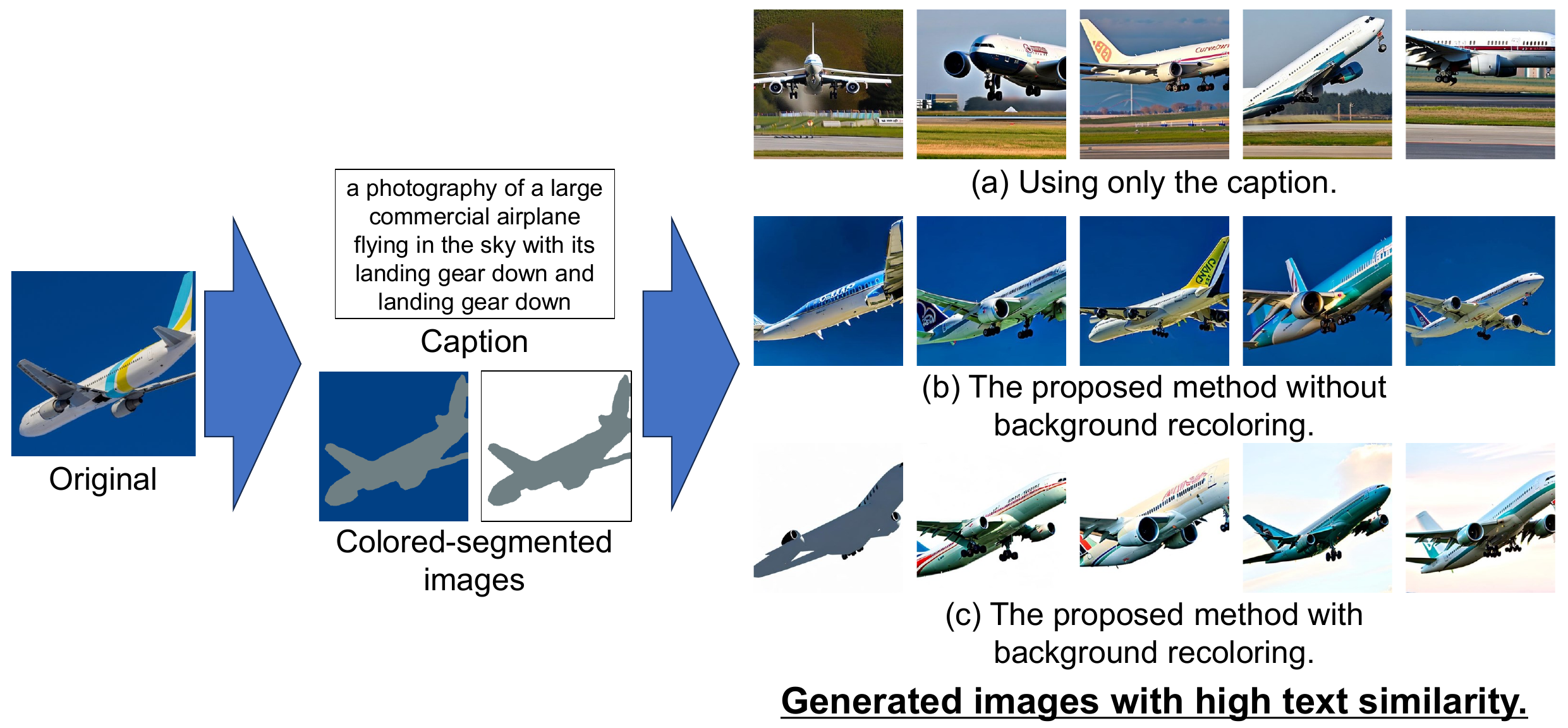}
  \caption{Top five images with the highest text similarity scores for each method for the original ``airplane'' image.
  The proposed method without background recoloring can reconstruct the entire color information of original image, including the background.
  In contrast, the proposed method with background recoloring results in the loss of background color information.}
  \label{fig:bert-score-rank-1}
\end{figure}

\subsection{Characteristics of Images Generated Using Each Method}
\label{sec:characteristirc}

Fig.~\ref{fig:bert-score-rank-1} shows the top five images with the highest text similarity scores generated using each method for one ``airplane'' image.
Note that the stop words were not removed from the captions in this result.
The images generated using only captions exhibit large composition and color information differences compared with the original image.
In contrast, the compositions of the images generated using the proposed methods are similar to those of the original image.
These results demonstrate that images with a composition similar to that of the original image can be generated by inputting the multi-modal semantic features contained in the original image into the image-generation model.
However, in Fig.~\ref{fig:bert-score-rank-1} (c), the background color information of the original image is lost in the generated images because background recoloring sets a color other than that of the target object to white.
This result suggests that in situations where background information is important, inputting all the color information, including that of the background, is advantageous.

\begin{figure}[b]
  \centering
  \includegraphics[clip,width=0.99\linewidth]{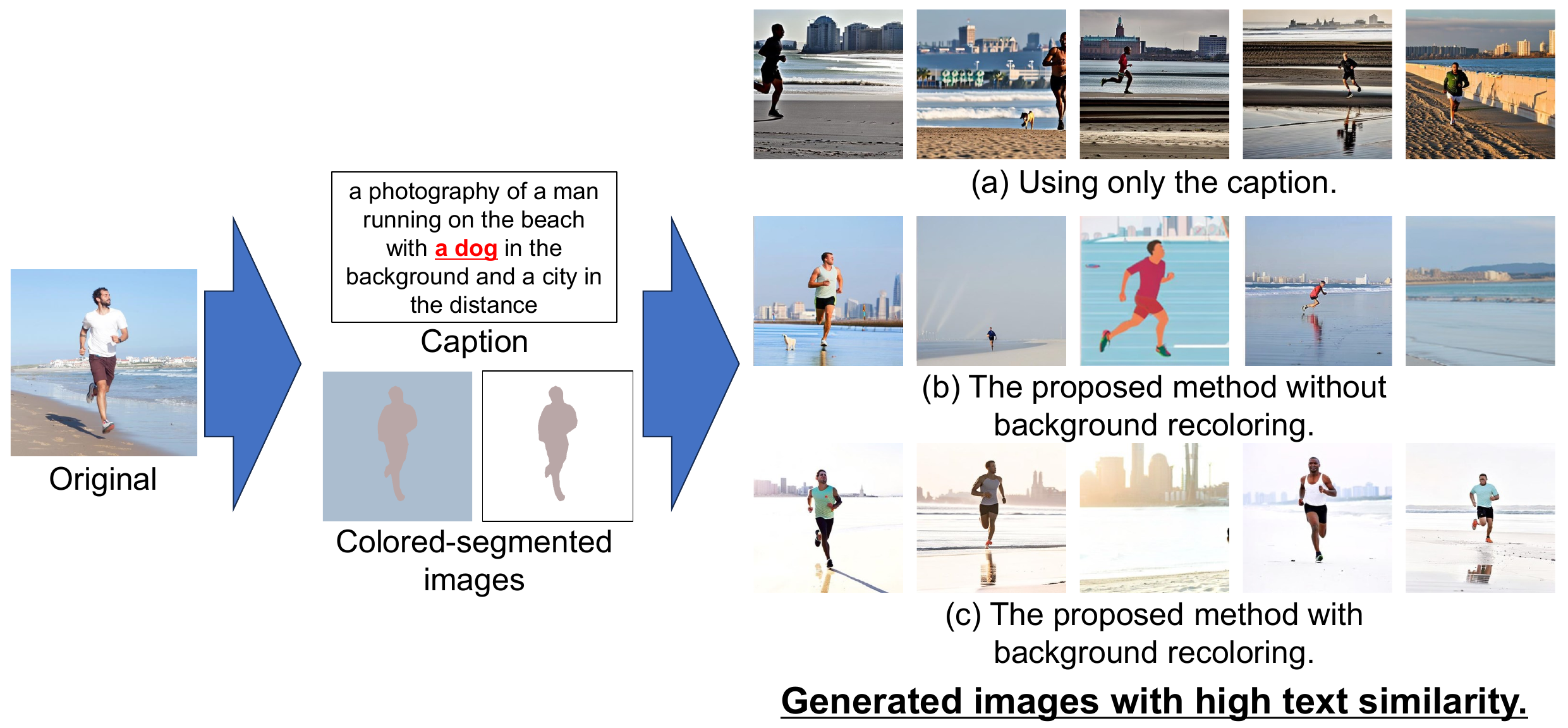}
  \caption{Top five images with the highest text similarity scores generated by each method for the original ``person'' image.
  The generated images with irrelevant objects exhibit high text similarity scores because the caption contains a word about the object that is not included in the original image.}
  \label{fig:bert-score-rank-2}
\end{figure}

\begin{figure}[b]
  \centering
  \includegraphics[clip,width=0.9\linewidth]{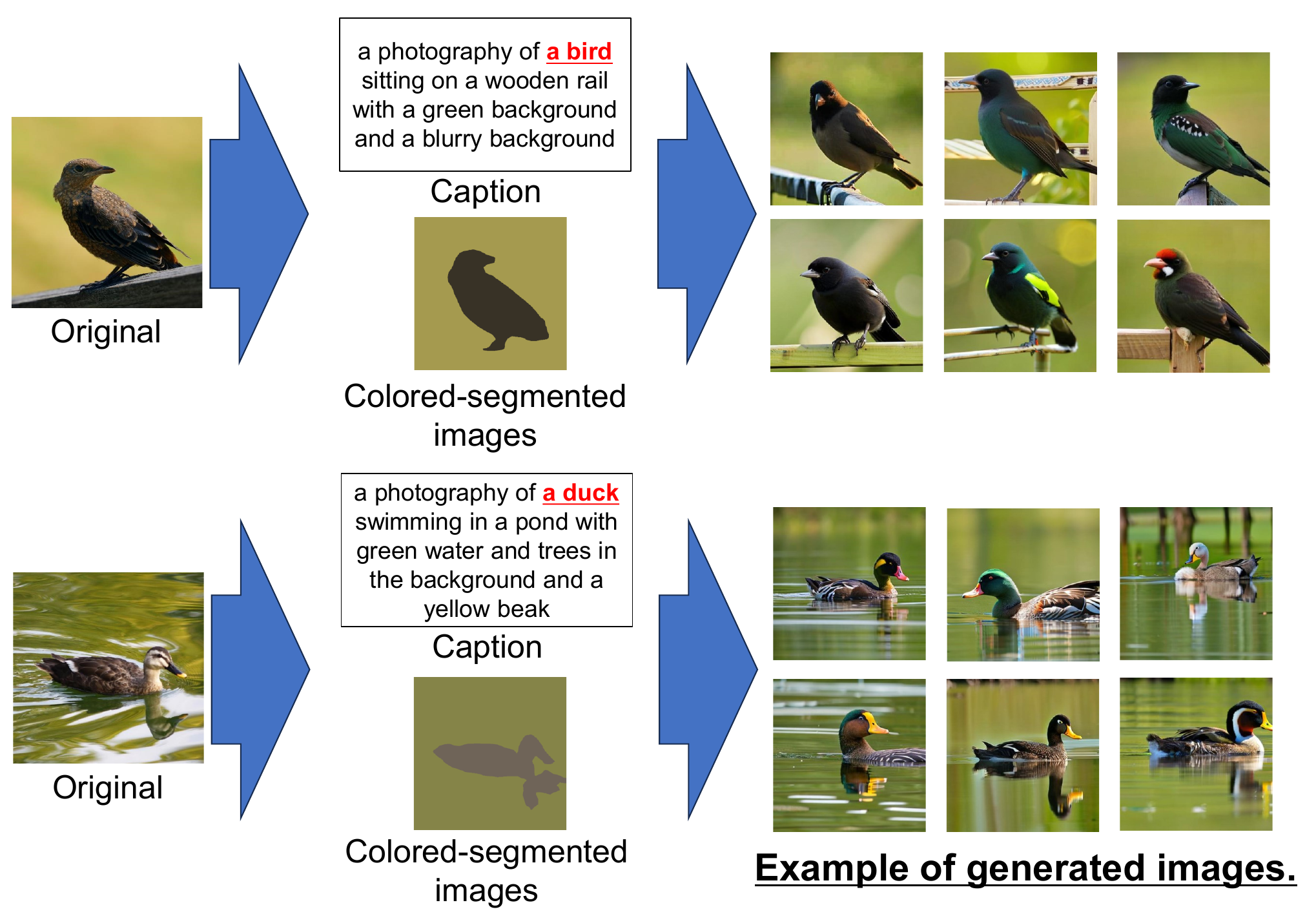}
  \caption{Semantic features and examples of the images generated using the ``bird'' images.
  If the caption contains specific words describing the target object, such as the type of living thing, this content is reflected in the generated images.}
  \label{fig:bert-score-example-bird}
\end{figure}

However, as shown in Fig.~\ref{fig:match-rate-rank-airplane-1} (b) and Fig.~\ref{fig:bert-score-rank-2} (b), when the colors of the background and target object are similar, the composition and position of the object in the generated images deviate from that in the original image.
This is because, if the RGB values of the object and background colors are close, the object boundary becomes ambiguous, and the image-generation model cannot recognize the position and composition of the object in the colored-segmented image.
In such situations, background recoloring that sets the background color to white is effective for making the object in the colored-segmented image stand out.

\begin{figure}[b]
  \centering
  \includegraphics[clip,width=0.9\linewidth]{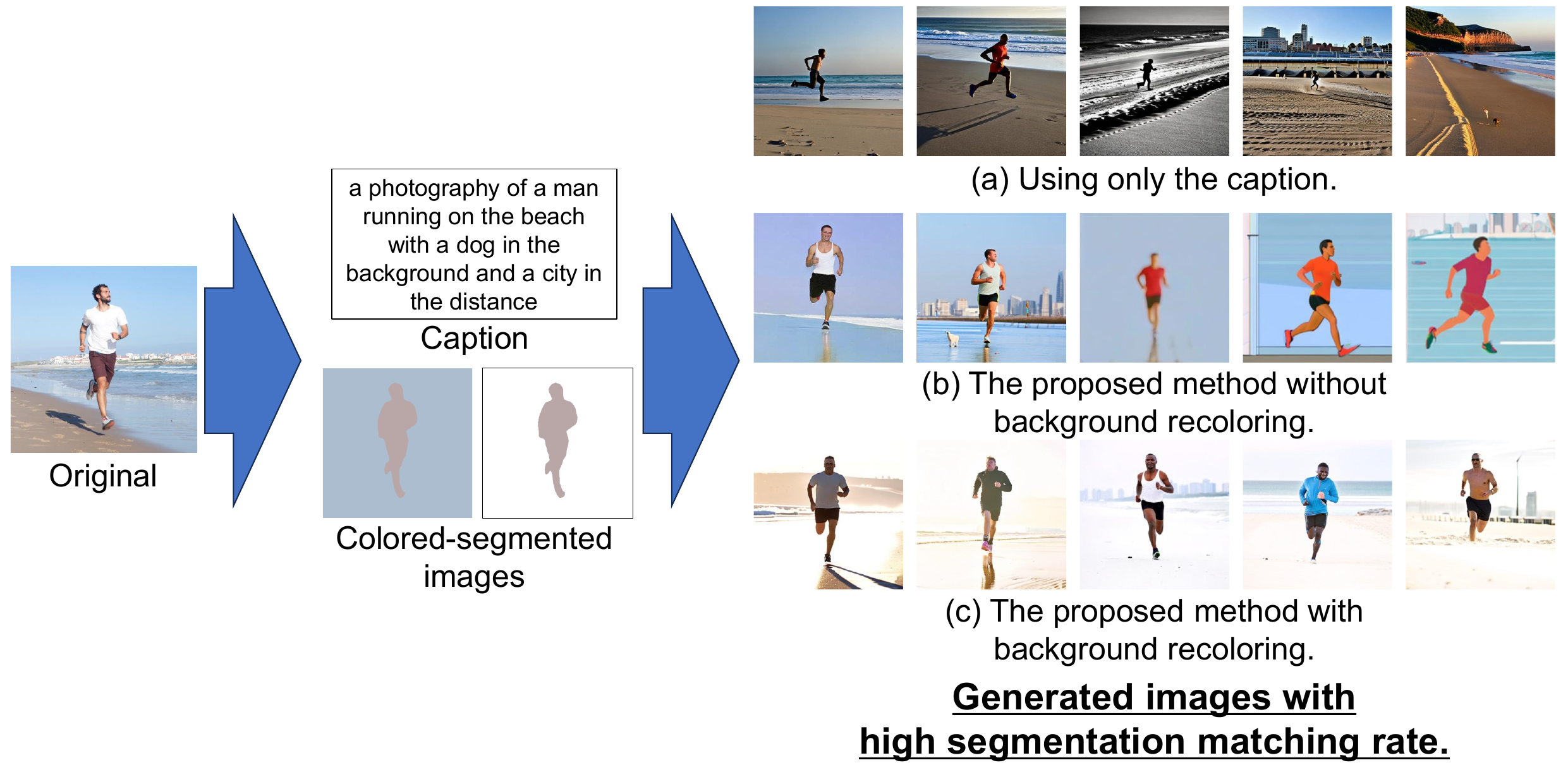}
  \caption{Top five images with the highest segmentation matching rates generated using each method for the original ``person'' image.
  The generated images with similar composition and position of the target object exhibit higher rates compared to the text similarity scores.}
  \label{fig:match-rate-rank-person}
\end{figure}

\begin{figure}[b]
  \centering
  \includegraphics[clip,width=0.99\linewidth]{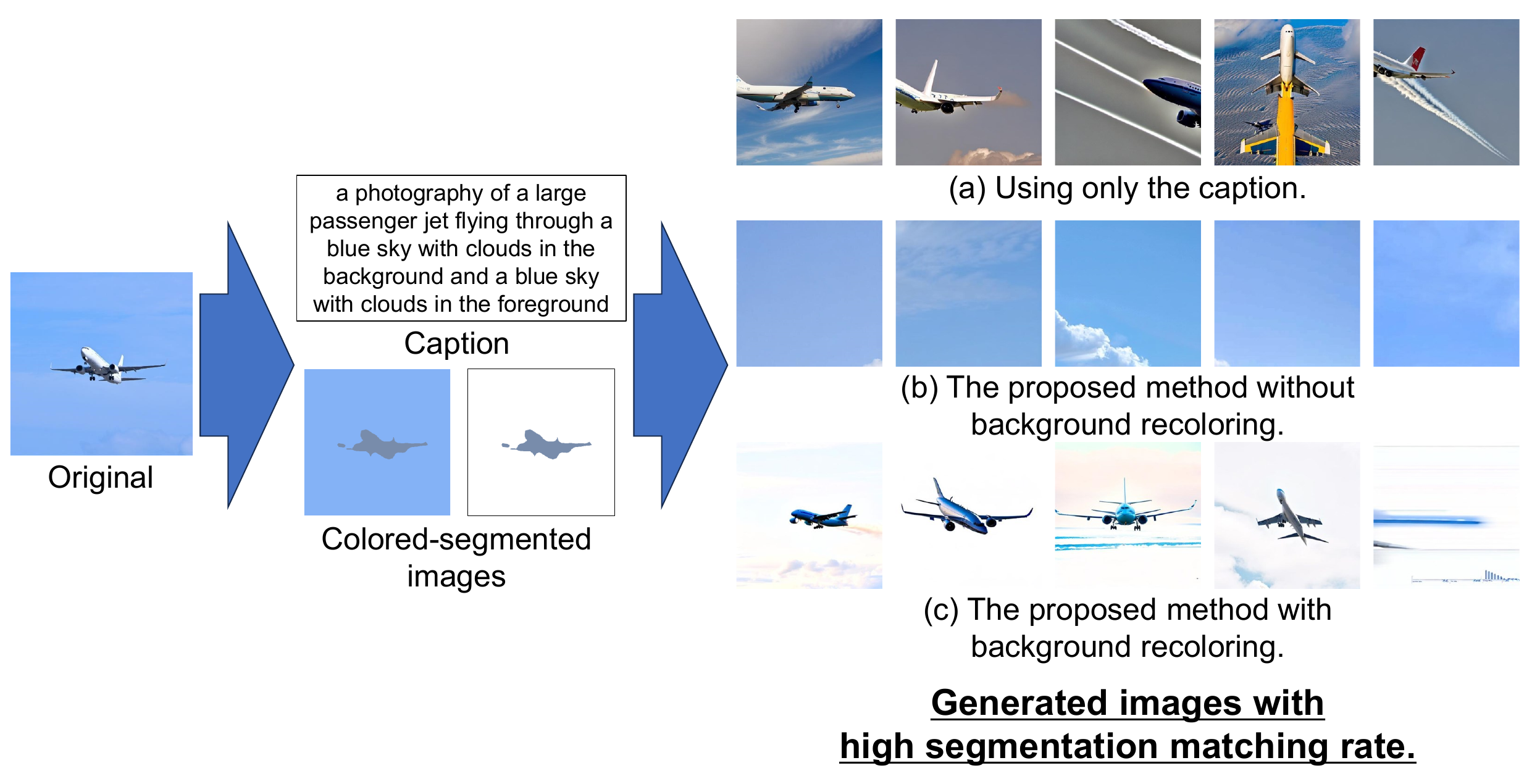}
  \caption{Top five images with the highest segmentation matching rates generated using each method for the original ``airplane'' image.
  The images with no objects generated using the proposed method without background recoloring exhibit high matching rates.
  Therefore, it is necessary to reconsider the calculation method for the segmentation matching rate in a future study.}
  \label{fig:match-rate-rank-airplane-2}
\end{figure}

\begin{figure*}[t]
  \centering
  \begin{tabular}{c}
    \begin{minipage}{0.32\hsize}
      \centering
      \includegraphics[clip, width=0.75\textwidth]{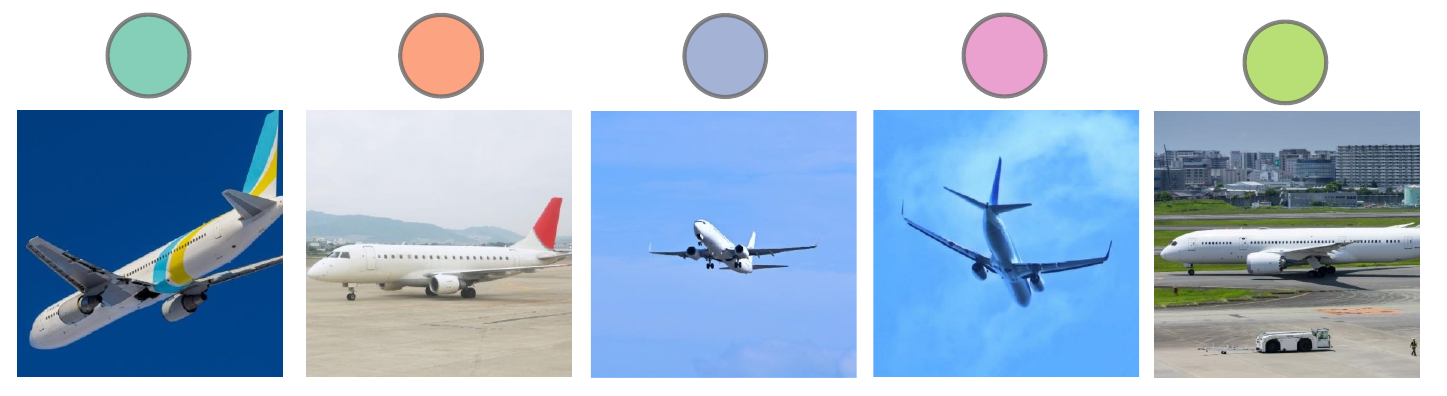}
      \includegraphics[clip, width=0.9\textwidth]{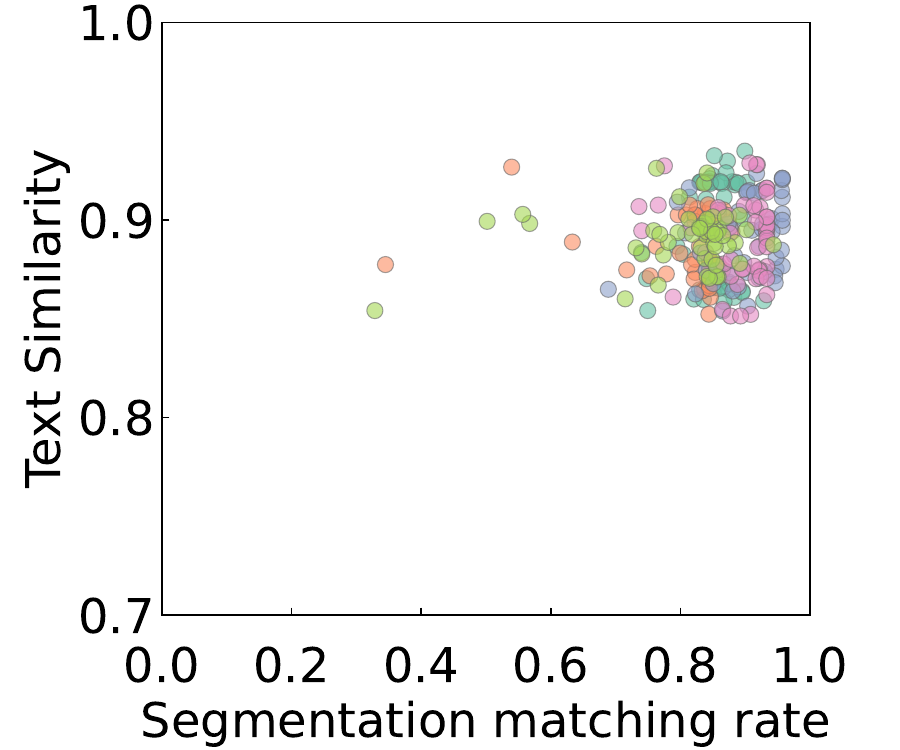}
    {\footnotesize
      \\ (a)~Airplane.}
    \end{minipage}
    \begin{minipage}{0.32\hsize}
      \centering
      \includegraphics[clip, width=0.75\textwidth]{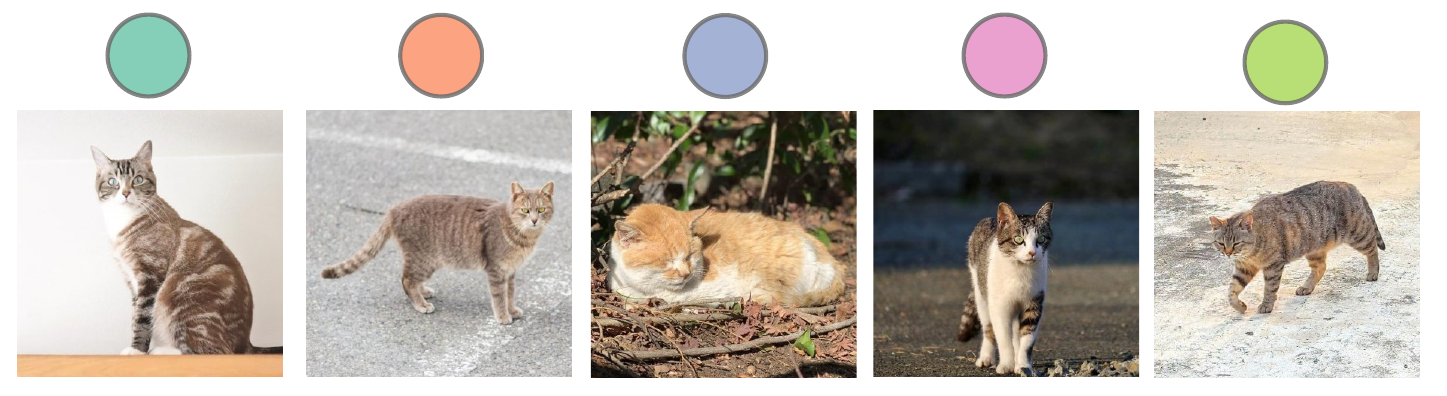}  
      \includegraphics[clip, width=0.9\textwidth]{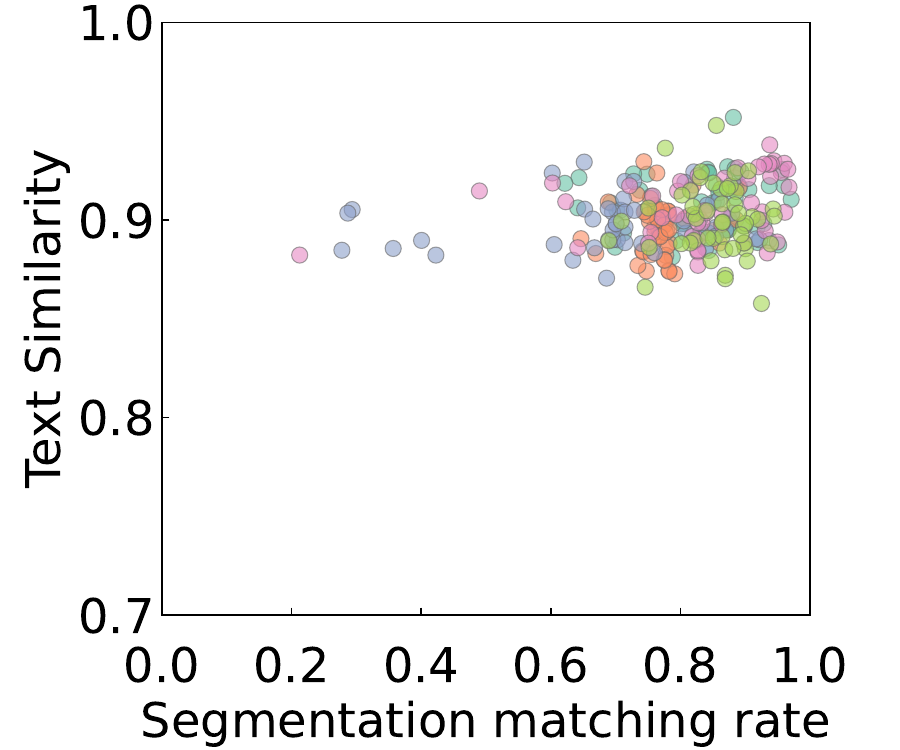}
    {\footnotesize
      \\ (b)~Cat.}
    \end{minipage}
    \begin{minipage}{0.32\hsize}
      \centering
      \includegraphics[clip, width=0.75\textwidth]{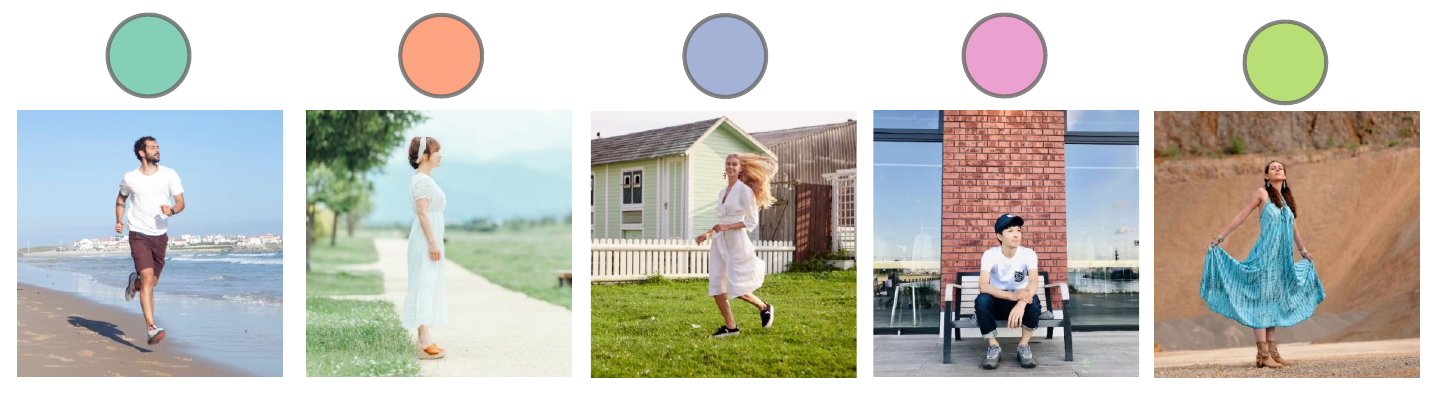} 
      \includegraphics[clip, width=0.9\textwidth]{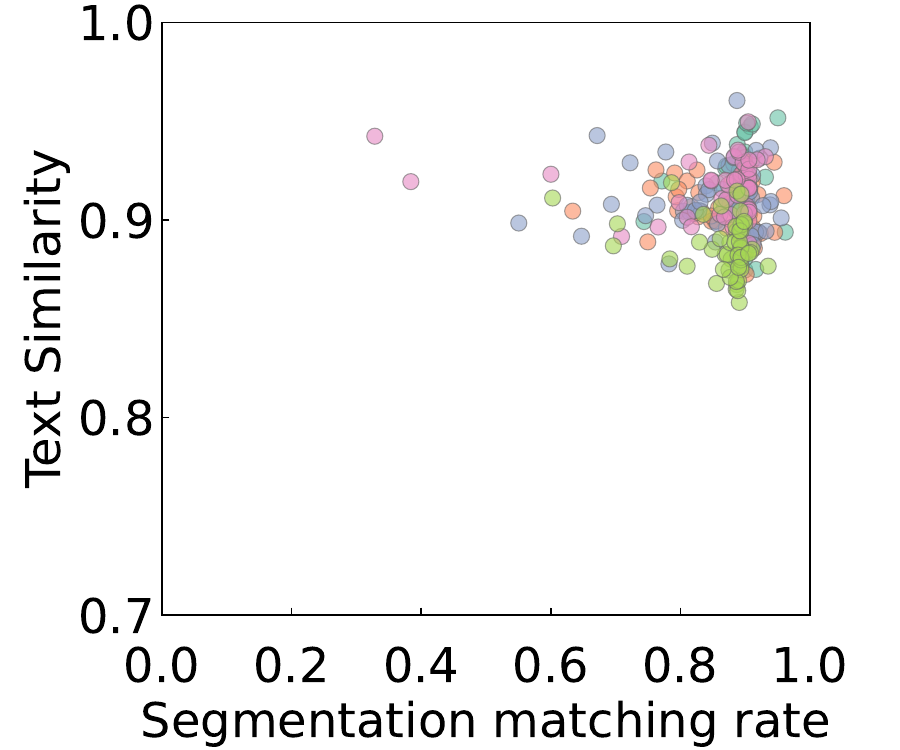}
    {\footnotesize
      \\ (c)~Person.}
    \end{minipage}
  \end{tabular}
  \caption{Scatter plots of text similarity scores and segmentation matching rates of the images generated using proposed method without background recoloring.
  To prevent outputting images with low scores, it is necessary to select the output result from multiple generated images based on their semantic similarity.}
  \label{fig:scatter}
\end{figure*}

\subsection{Impacts of Captions on Contents of Generated Images}
\label{sec:caption}

Fig.~\ref{fig:bert-score-rank-2} shows the top five images with the highest text similarity scores generated using each method for one ``person'' image.
In this figure, the caption generated from the original image contains an unrelated word ``dog.''
This phenomenon is caused due to misidentification by the image captioning algorithm.
Although there were no dogs in the original image, some of the generated images contained dog-like objects, and these images exhibited high text similarity scores.
To prevent this phenomenon, further improvements in the caption generation algorithm are required.

Fig.~\ref{fig:bert-score-example-bird} shows the semantic features generated from the two ``bird'' images and examples of images generated using these features.
These results indicate that if the original image contains a living thing, the generated caption may contain a description of a specific type of this living thing.
For example, in the caption generated from the image shown at the top of Fig.~\ref{fig:bert-score-example-bird}, the target object is simply described as a ``a bird.''
By contrast, the caption generated from the image at the bottom of Fig.~\ref{fig:bert-score-example-bird} describes the object using the specific species name ``a duck.''
Consequently, the images generated using the bottom caption contain duck-like birds.
This result suggests that the content of the generated images is affected by the granularity of the descriptive information.
Therefore, it can be possible to adjust the granularity of the descriptive information included in the caption based on the situation and application.

\subsection{Advantages and Disadvantages of Evaluating Text Similarity and Segmentation Matching Rate}
\label{sec:scoring-advantage}

Fig.~\ref{fig:bert-score-rank-2} and Fig.~\ref{fig:match-rate-rank-person} show the top five generated images with the highest text similarity scores and segmentation matching rates generated by each method.
Note that the images in these figures are generated based on the same original ``person'' image.
As shown in Fig.~\ref{fig:bert-score-rank-2}, because the caption contains the word ``dog,'' the generated images that contain dog-like objects exhibit high text similarity scores.
Additionally, because the captions include the words ``beach'' and ``city,'' the generated images containing these elements in the background also exhibit high text similarity scores.
These results indicate that the scoring procedure for text similarity is effective for evaluating the type of target object and broad background information.

As shown in Fig.~\ref{fig:match-rate-rank-person}, the generated images with high segmentation matching rates are more similar to the original image in terms of the composition and position of the target object than those with high text similarity scores.
Therefore, these results indicate that the segmentation matching rate is effective for evaluating the composition and position of the target object in the generated images.

Fig.~\ref{fig:match-rate-rank-airplane-2} shows the top five generated images with the highest segmentation matching rates for the ``airplane'' image by each method.
As shown in Fig.~\ref{fig:match-rate-rank-airplane-2} (b), if the target object in the original image is too small, the images generated without the target object exhibit higher scores than the other images.
This is because the number of pixels labeled ``airplane'' is considerably lower than the number of pixels; therefore, the matching rate of the images generated without any object where all pixels are labeled ``background'' is higher.
To address this issue, future studies should develop a novel calculation method for the segmentation matching rate.

\subsection{Effectiveness of the Proposed Methods and Scoring Procedures}
\label{sec:effectiveness}

Fig.~\ref{fig:scatter} shows the scatter plots of the text similarity scores and segmentation matching rates for the images generated by the proposed method without background recoloring.
Note that the stop words were not removed in the text similarity evaluations, and the color of each point indicates the type of semantic feature in the original image used for image generation.
Some of the generated images exhibit significantly lower scores across both metrics than the other images.
This implies that if only a single image is reconstructed as an output in an image-generation-based transmission, the image can be semantically different from the original image.
Therefore, as employed in the proposed method, it is essential to select an image that is similar to the original image from multiple generated images based on the proposed scoring procedure.

Based on these results, the proposed method realizes both data reduction and image generation using semantic information that is more similar to the original image than using only a single semantic feature of the image.
In addition, these results indicate that the text similarity scores and segmentation matching rates can quantify the semantic similarity of the generated images in terms of descriptive and segment information, respectively.
The proposed method without background recoloring can realize image transmission while maintaining the background color information around the target object.
However, if the colors of the target object and the background are similar, the image-generation model cannot correctly recognize the segment information of the colored-segmented image.
This issue can be addressed by applying background recoloring that changes the colors of pixels labeled ``background'' to white.

\section{Conclusion}
\label{sec:conclusion}

This paper proposed an image-generation-based transmission method.
In the proposed method, a transmitter extracts multi-modal semantic features from an image and transmits only them to a receiver.
The receiver generates multiple images using an image-generation model and selects an output image based on the semantic similarity between the original and generated images.
Therefore, the proposed method can realize both a significant reduction in the transmitted data size and reconstruction of the image with the semantic information required by the receiver.
The evaluation results indicates that the proposed method can reduce the data size compared with typical image compression algorithms.

In the proposed method, the receiver must evaluate the similarity between the original and generated images using only the received data to select the output.
Therefore, this study proposed two scoring methods for comparing them between the original and generated images, and an experiment was conducted to verify the effectiveness of the proposed transmission method and scoring procedures.
The results indicated that the proposed method using multi-modal semantic features can generate images that are more similar to the original images than those using only a single feature.
In addition, the proposed scoring procedures can quantify the semantic similarity between images.
The text similarity scores can compare the types of objects and rough background information in images and the segmentation matching rates can compare the composition and position of objects in the images.
However, we found that if the target object in the image was too small, the generated images with no objects exhibited high segmentation matching rates.

In a future work, we will improve the scoring procedures used for evaluating the semantic similarity.
In the proposed scoring procedure for the segmentation matching rate, generated images with no objects had a high score when the target object in the original image was too small.
This is because the matching rate is calculates based on the number of pixels including those labeled as ``background.''
Therefore, it is necessary to develop a calculation method that considers only the pixels labeled as ``target object.''
Moreover, we plan to develop an image reconstruction method based on the evaluation results of the semantic similarity.
In the proposed method, the receiver selects the output from the generated images based on semantic similarity.
However, if all generated images have low similarity scores, the receiver cannot select the result required by its application.
To address this issue, it is necessary to develop a method to regenerate images or, if necessary, request the sender to transmit the semantic features again.
Finally, we plan to apply the proposed transmission method to video streaming.
The current proposed method does not focus on a processing delay caused by extraction of semantic features and image generation. However, this issue may be addressed by employing faster machine learning algorithms for real-time applications~\cite{bib:stream-diffusion,bib:fast-semantic-segmentation,bib:fast-image-captioning}.
If the proposed transmission method can be applied to video streaming applications, this method will achieve a higher compression rate than conventional video compression algorithms.

\section*{Acknowledgments}
This work was partly supported by NICT (JPJ012368C01101).

\bibliographystyle{ieicetr}
\bibliography{reference}

\profile[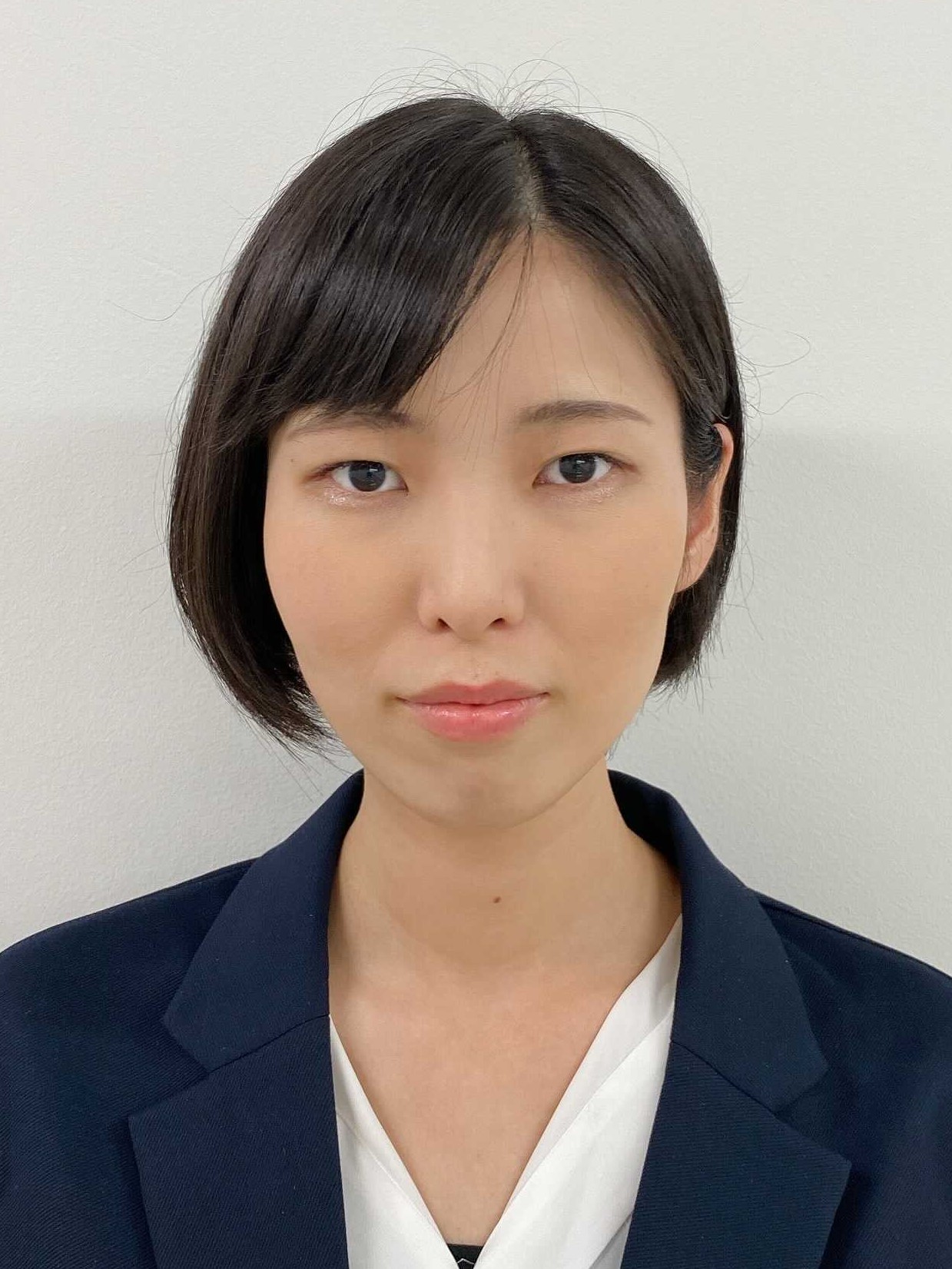]{Eri Hosonuma}
{received the B.E. and M.S. degrees in electronic information systems from Shibaura Institute of Technology, Tokyo, Japan, in 2020 and 2022, respectively. She is presently a doctoral course student at the Department of Socio-Cultural Environmental Studies, The University of Tokyo, Tokyo, Japan.}

\profile[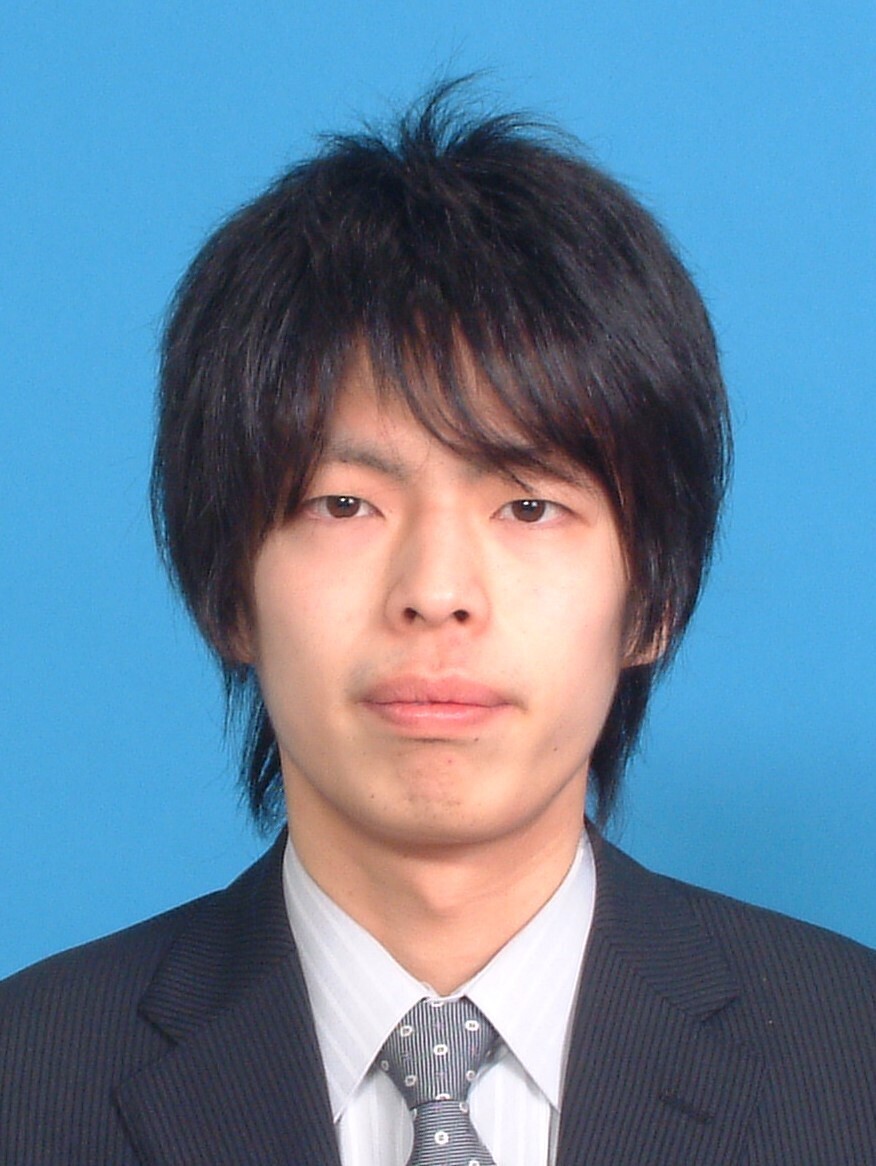]{Taku Yamazaki}
{received the B.E. and M.S. degrees in electronic information systems from Shibaura Institute of Technology, Tokyo, Japan, in 2012 and 2014, respectively. He received the D.E. degree in computer science and communications engineering from Waseda University, Tokyo, Japan, in 2017. He is presently an associate professor at Department of Electronic Information Systems, College of Systems Engineering and Science, Shibaura Institute of Technology, Saitama, Japan. His research interests include wireless networks, internet of things, and network security.}

\profile[BIO/Author-miyoshi.JPG]{Takumi Miyoshi}
{received his B.Eng., M.Eng., and Ph.D.\ degrees in electronic engineering from the University of Tokyo, Japan, in 1994, 1996, and 1999, respectively.  He started his career as a research associate in Waseda University from 1999 to 2001, and is presently a professor at Department of Electronic Information Systems, College of Systems Engineering and Science, Shibaura Institute of Technology, Saitama, Japan.  He is also a research fellow in Institute of Industrial Science, the University of Tokyo, Tokyo, Japan.  He was a visiting scholar in Laboratoire d'Informatique de Paris 6 (LIP6), Sorbonne Universit\'e, Paris, France, from 2010 to 2011.  His research interests include overlay networks, location-based services, and mobile ad hoc and sensor networks.}

\profile[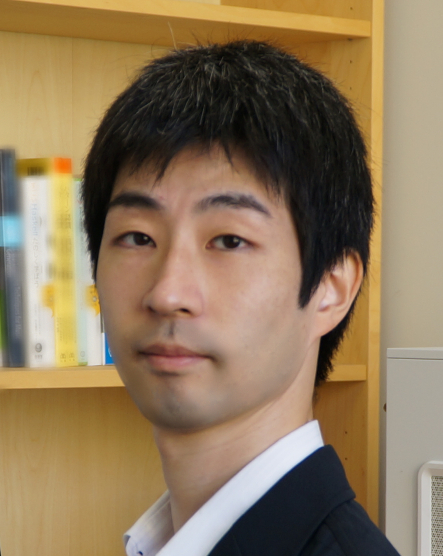]{Akihito Taya}
{received the B.E. degree in electrical and electronic engineering from Kyoto University, Kyoto, Japan in 2011, and the master and Ph.D. degree in Informatics from Kyoto University in 2013 and 2019, respectively. From 2013 to 2017, he joined Hitachi, Ltd., where he participated in the development of computer clusters. From 2019 to 2022, he was an assistant professor of Aoyama Gakuin University. He has been an assistant professor of The University of Tokyo, since 2022. He received the IEEE VTS Japan Young Researcher's Encouragement Award and the IEICE Young Researcher's Award in 2012 and 2018, respectively. His current research interests include distributed machine learning and human activity and emotion recognition using sensor networks. He is a member of the IEEE, ACM, IEICE and IPSJ.}

\newpage

\profile[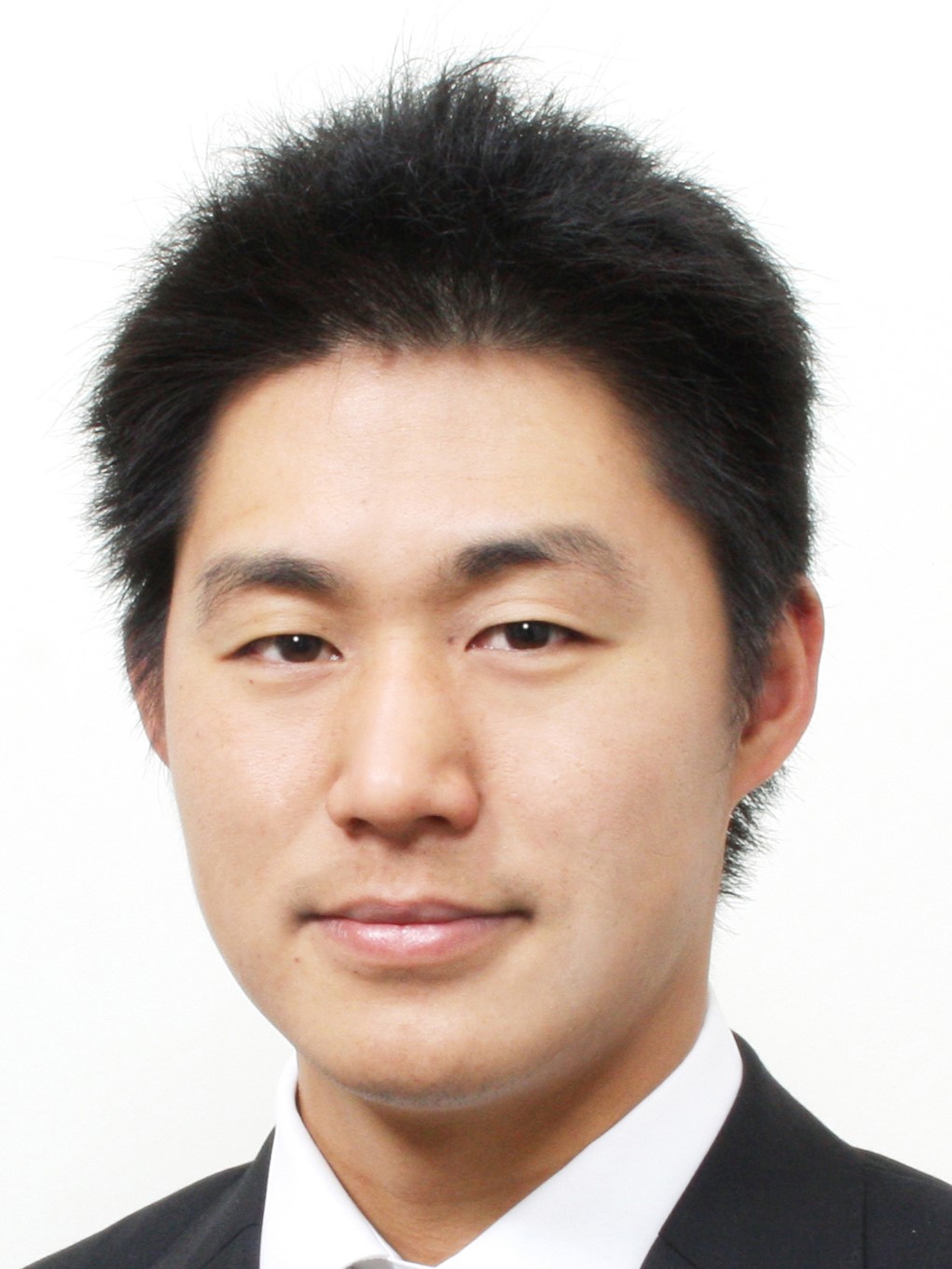]{Yuuki Nishiyama}
{is an Assistant Professor at the Center for Spatial Information Science in the University of Tokyo. He obtained M.S.(2014) in Media and Governance from Keio University, and Ph.D. in Media and Governance (2017) from Keio University, respectively. He had worked at Keio University in Japan and the University of Oulu in Finland, as a post-doctoral researcher respectively. He started work at the Institute of Industrial Science in the University of Tokyo as a Research Associate in 2019, and has held his current position since 2022. His current research interests include ubiquitous computing, contex-aware systems, human behavior change, and human ability augmentation. He is a member of ACM, IEEE, and Information Processing Society of Japan (IPSJ).}

\profile[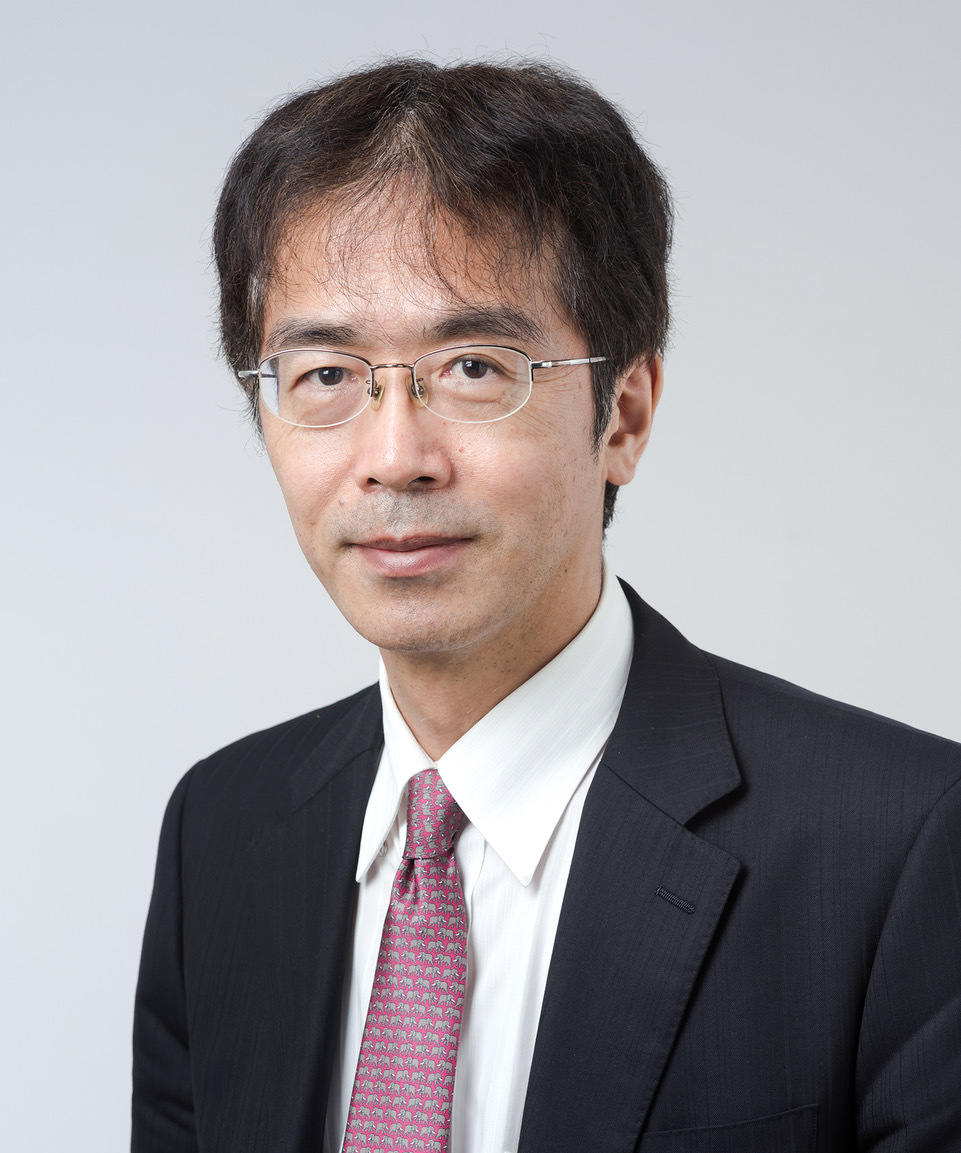]{Kaoru Sezaki}
{is the director of Center for Spatial Information Science at the University of Tokyo and co-appointed as professor of Institute of Industrial Science, the University of Tokyo. He is a steering member of e-Health Technical Committee COMSOC. He has been general chair and TPC Chair of many IEEE international conferences. He also served as Treasurer of IEEE Tokyo Section as well as that of Japan Council from 2003 to 2004.He received B.Eng., M.Eng., and Ph.D. degrees from the University of Tokyo, Tokyo, Japan, in 1984, 1986, and 1989, respectively, all in Electrical Engineering. Since 1989, he has been with the University of Tokyo. He was a Visiting Researcher at University of California at San Diego in 1996. His research interests include e-Health, sensor networks, IoT, and urban computing.}

\end{document}